\pdfoutput=1
\documentclass[12pt,preprint]{aastex}
\usepackage{graphicx}

\newcommand{\et}{et al.}
\newcommand{\kms}{km s$^{-1}$}
\newcommand{\ha}{H$\alpha$}
\newcommand{\solar}{\ifmmode_{\sun}\;\else$_{\sun}\;$\fi}

\newcommand{\HII}{H$\,${\sc ii}$\,$}
\newcommand{\HI}{H$\,${\sc i}$\,$}

\newcommand{\x}{\enspace}
\newcommand{\xx}{\enspace\enspace}

\newcommand{\sigcrit}{$\Sigma_{c}$}
\newcommand{\siggas}{$\Sigma_g$}

\begin{document}

\title{The outer disks of dwarf irregular galaxies}

\author{Deidre A. Hunter\footnote{Visiting Astronomer, Kitt Peak National Observatory, National Optical Astronomy Observatory, which is operated by the Association of Universities for Research in Astronomy (AURA) under cooperative agreement with the National Science Foundation.}~\footnote{Guest Observer, Very Large Array,
National Radio Astronomy Observatory.
The National Radio Astronomy Observatory is a facility of the National Science Foundation operated under cooperative agreement by Associated Universities, Inc.}
}
\affil{Lowell Observatory, 1400 West Mars Hill Road, Flagstaff, Arizona
86001 USA}
\email{dah@lowell.edu}

\author{Bruce G. Elmegreen}
\affil{IBM T. J. Watson Research Center, PO Box 218, Yorktown Heights,
New York 10598 USA}
\email{bge@watson.ibm.com}

\author{Se-Heon Oh\footnote{Current address: International Centre for Radio Astronomy Research,
The University of Western Australia, M468, 35 Stirling Highway, Crawley, WA 6009 AUSTRALIA}
}
\affil{Department of Astronomy, University of Cape Town, Private Bag X3, Rondebosch 7701, Republic of
South Africa}
\email{seheon-oh@ast.uct.ac.za}

\author{Ed Anderson$^*$}
\affil{Department of Physics and Astronomy, Northern Arizona University,
P.O. Box 6010, Flagstaff, AZ 86011-6010 USA}
\email{Ed.Anderson@nau.edu}

\author{Tyler E. Nordgren}
\affil{Department of Physics, University of Redlands, 1200 East Colton Avenue,
Redlands, CA 92373 USA}
\email{tyler\_nordgren@redlands.edu}

\author{Philip Massey$^*$,
Nick Wilsey\footnote{Current address: Truman State University, 100 E Normal, Kirksville, Missouri USA 63501},
and Malanka Riabokin\footnote{Current address: Department of Physics and Astronomy,
Michigan State University, East Lansing, Michigan 48824-2320 USA}}
\affil{Lowell Observatory, 1400 West Mars Hill Road, Flagstaff, Arizona 86001 USA}
\email{phil.massey@lowell.edu, njw459@truman.edu, riabokin@msu.edu}

\begin{abstract}
In order to explore the properties of extreme outer stellar disks, we obtained
ultra-deep $V$ and {\it GALEX} UV images of 4 dwarf irregular galaxies and one Blue Compact Dwarf galaxy
and ultra-deep $B$ images of 3 of these.
Our $V$-band surface photometry extends to 29.5 magnitudes arcsec$^{-2}$.
We convert the FUV and $V$-band photometry, along with \ha\ photometry obtained in a larger survey,
into radial star formation rate profiles
that are sensitive to timescales from 10 Myrs to the lifetime of the galaxy.
We also obtained \HI-line emission data and compare
the stellar distributions, surface brightness profiles, and star formation rate profiles to \HI-line emission maps,
gas surface density profiles, and gas kinematics. Our data lead us to two general observations:
First, the exponential disks in these irregular galaxies are extraordinarily regular.
We observe that the stellar disks continue to decline exponentially as far as our measurements extend.
In spite of lumpiness in the distribution of young stars and
\HI distributions and kinematics that have significant unordered motions, sporadic processes that have built the
disks---star formation, radial movement of stars, and perhaps even perturbations from the outside---have,
nevertheless, conspired to produce standard disk profiles.
Second, there is a remarkable continuity of star formation throughout these disks over time.
In four out of five of our galaxies the star formation rate in the outer disk measured from the FUV tracks that
determined from the $V$-band, to within factors of 5,
requiring star formation at a fairly steady rate over the galaxy's lifetime.
Yet, the \HI surface density profiles generally decline with radius more shallowly than the
stellar light, and the gas is marginally gravitationally stable against collapse into clouds.
Outer stellar disks are challenging our concepts of star formation and disk growth
and provide a critical environment in which to understand processes that mold galaxy disks.
\end{abstract}


\section{Introduction} \label{sec-intro}

Outer parts of galaxy disks present an extreme environment for star formation.
Dwarf irregular (dIm) galaxies already challenge models of star formation
because of their low gas densities even in the central regions
(Hunter \& Plummer 1996; Meurer \et\ 1996; van Zee \et\ 1997; Hunter,
Elmegreen, \& Baker 1998).
Outer parts of dIm galaxies with their exceptionally low gas densities, therefore, present a particularly difficult
test of our understanding of the cloud/star formation process.
Furthermore, knowledge of the star formation histories of outer stellar disks provides
information on the way in which the disk has grown with time and tests
current models for disk growth (see, for example, Azzollini \et\ 2008).

Stars have formed even to very low disk surface brightness levels.
For example, Saha \et\ (2010) have detected an exponential stellar disk in the LMC
that extends to 12 disk scale lengths, the equivalent of an $I$-band surface brightness of 34 mag arcsec$^{-2}$.
Although the number of young stars relative to older stars declines with radius,
stars as young as a few Gyr are found to about 8.3 disk scale lengths.
In the late-type spiral M33, Barker \et\ (2011) have found stars younger than 0.5 Gyr at 4 and 5 disk scale lengths,
although the bulk of the stars are older.
And in M31, Richardson \et\ (2008) find evidence for star formation over the past 0.25-2 Gyr as far out as 56 kpc.
Furthermore, satellite ultraviolet imaging has revealed extended ultraviolet disks in many spiral galaxies
that indicate on-going star formation in outer disks
(Thilker \et\ 2007).
For example, {\it GALEX} images of M83 revealed UV-bright stellar complexes to 7 kpc, about 4 times the radius
at which most of the \HII\ regions were detected (Thilker \et\ 2005).

Given the importance of outer stellar disks to our understanding of galaxy formation and
evolution, we have conducted a program to
detect and measure the extreme outer stellar disks of a sample of
dIm galaxies through ultra-deep imaging in the optical and ultraviolet.
In addition we have mapped the galaxies in \HI-line emission.
We use these data to examine the structure,
stellar populations, and star formation activity in extreme outer dIm disks.

Our program objects were selected from a large survey of
dIm, Blue Compact Dwarf (BCD), and Sm galaxies (Hunter \& Elmegreen 2006).
They were chosen to be representative of the larger dIm sample
in terms of absolute magnitude $M_V$ and star formation rate.
They were also chosen to be
relatively free of nearby bright stars that would compromise
photometry at ultra-low light levels.
The new deep imaging data will be referred to here as the ``deep'' data,
while the shallower survey data of the larger sample will be referred
to as ``survey'' data.
The galaxies and several important characteristics are
listed in Table \ref{tab-sample}.

\section {Observations} \label{sec-obs}

\subsection{Optical images}

\subsubsection{Data acquisition and basic reductions}

\begin{figure}
\epsscale{0.9}
\includegraphics[angle=0,width=1.0\textwidth]{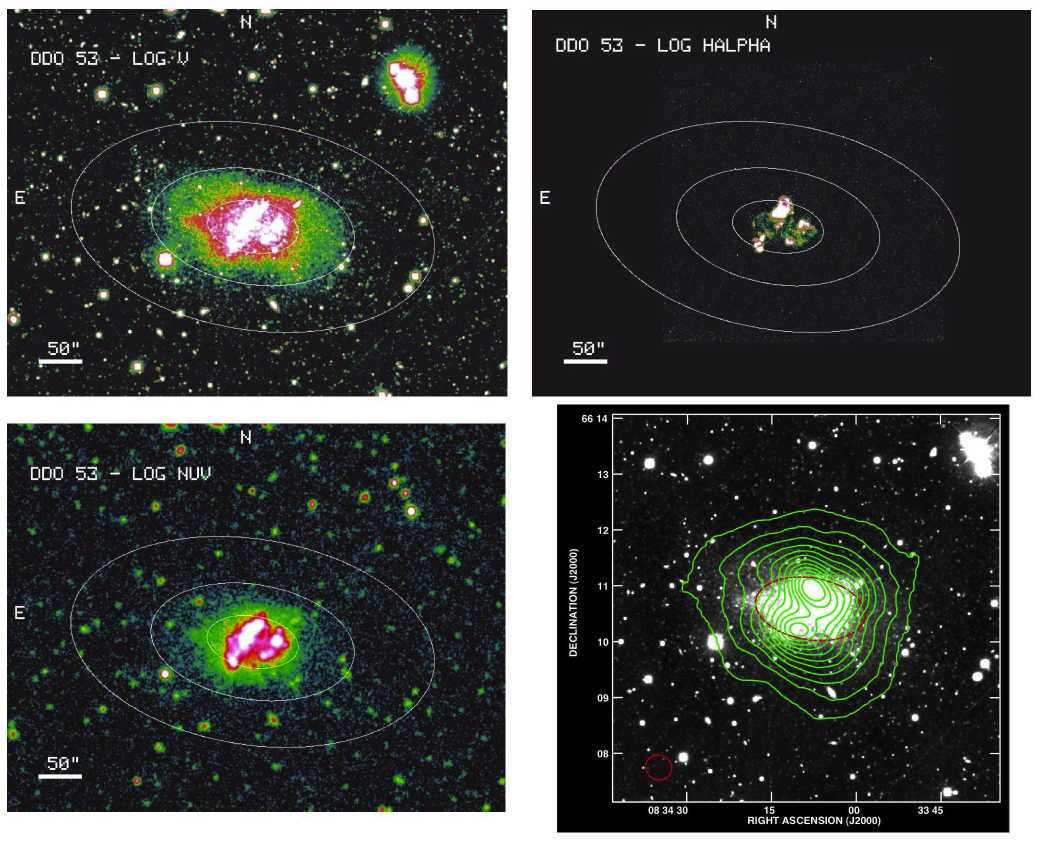}
\caption{{\it Left and Upper Right:} $V$, H$\alpha$, and NUV images of DDO 53,
shown as the logarithm in order to show a wider range of brightness levels.
The white ellipses show the breaks in the surface photometry and the outer-most ellipse
that was used in the surface photometry.
{\it Bottom Right:}
$V$-band image with contours of integrated \HI\ emission superposed:
 1.9 to 64.3 in steps of 5.67$\times10^{20}$ cm$^{-2}$.
The large red ellipse denotes the radius of the major
break in the $V$-band surface photometry.
The small red ellipse in the lower left corner shows the FWHM of the VLA beam.
\label{fig-d53}}
\end{figure}

\begin{figure}
\epsscale{0.9}
\includegraphics[angle=0,width=1.0\textwidth]{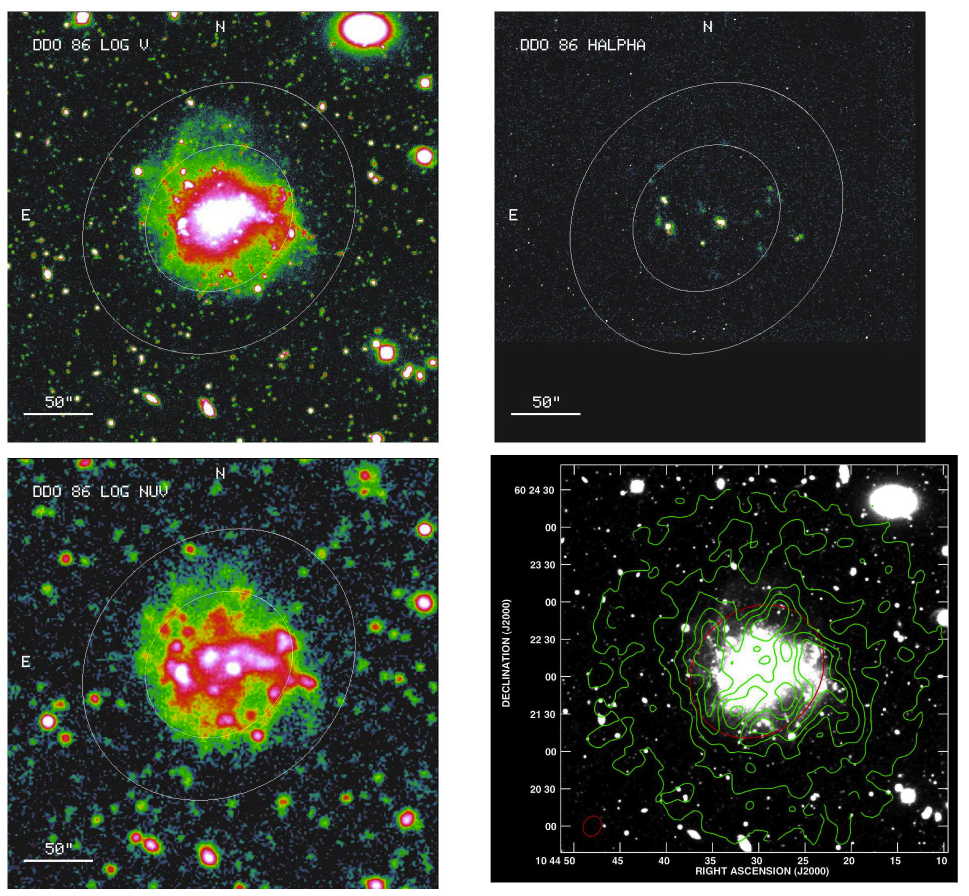}
\caption{
As in Figure 1, for DDO 86.
The H$\alpha$ image is not shown as the logarithm.
For DDO 86, photometry was measured at the position angle defined by the moment zero map fit to the \protect\HI\ kinematics.
Contours of integrated \HI\ emission are
13.5 to 22.9 in steps of 5.38$\times10^{20}$ cm$^{-2}$.
\label{fig-d86}}
\end{figure}

\begin{figure}
\epsscale{0.9}
\includegraphics[angle=0,width=1.0\textwidth]{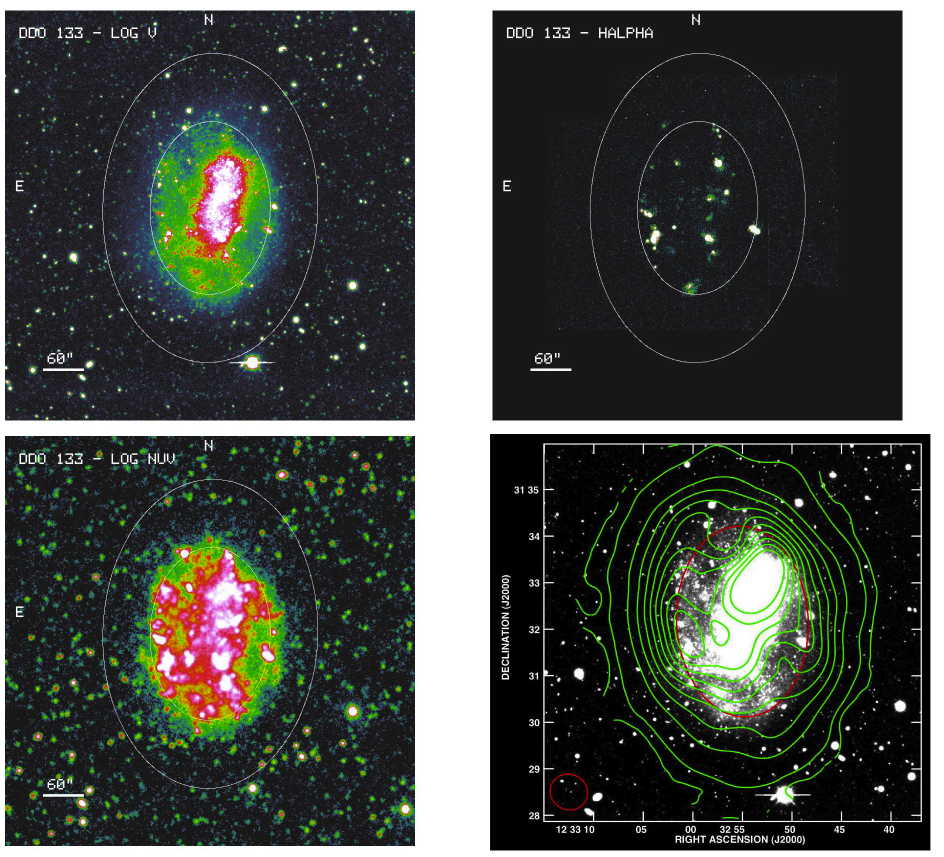}
\caption{
As in Figure 1, for DDO 133.
The H$\alpha$ image is not shown as the logarithm.
Contours of integrated \HI\ emission are
0.06, 0.3, 1.58 to 11.7 in steps of 1.26$\times10^{20}$ cm$^{-2}$.
\label{fig-d133}}
\end{figure}

\begin{figure}
\epsscale{0.9}
\includegraphics[angle=0,width=1.0\textwidth]{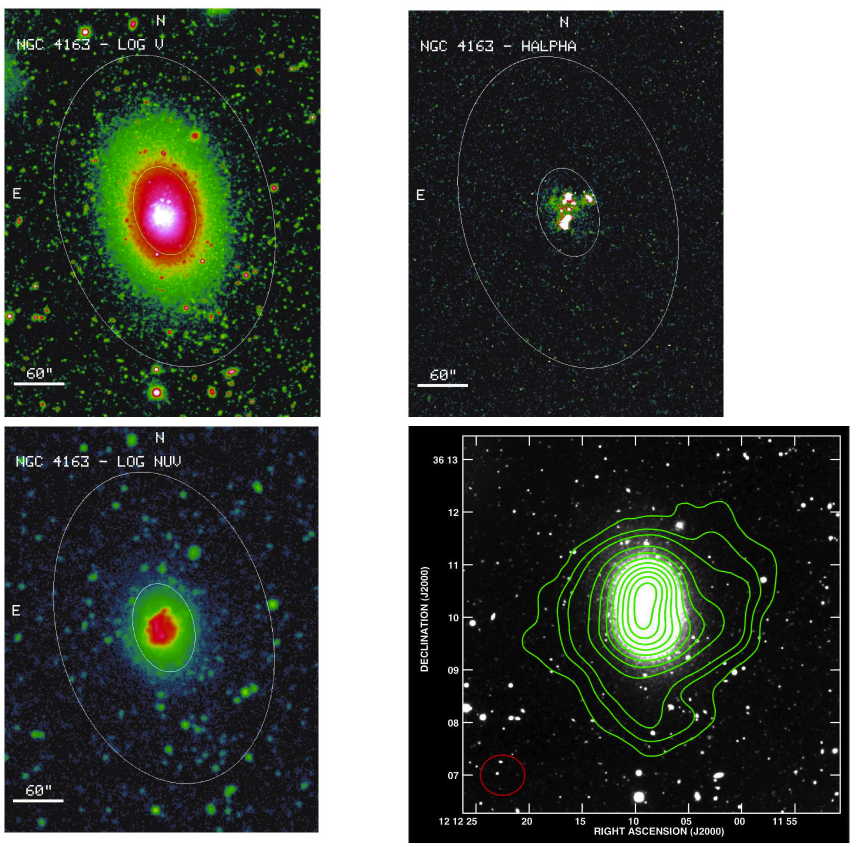}
\caption{
As in Figure 1, for NGC 4163.
The H$\alpha$ image is not shown as the logarithm.
Contours of integrated \HI\ emission are
1.8 and 3.03 to 7.89 in steps of 2.43$\times10^{19}$ cm$^{-2}$
and 1.7 to 8.1 in steps of 0.91$\times10^{20}$ cm$^{-2}$.
\label{fig-n4163}}
\end{figure}

The deep optical imaging observations are summarized in
Table \ref{tab-obs} and $V$-band and \ha\
images are shown in Figures \ref{fig-d53} through \ref{fig-izw115}.
The \ha\ images are part of the survey data
(Hunter \& Elmegreen 2004).
Reductions were done using the {\it Image Reduction and Analysis Facility} (IRAF\footnote{IRAF is distributed by the
National Optical Astronomy Observatory, which is operated by the Association of Universities for Research in Astronomy
(AURA) under cooperative agreement with the National Science Foundation.}).

In May 2002 we 
obtained 
deep $B$ and $V$ images of DDO 133 and $V$ images of IZw 115
with the Mosaic Imager on the
Kitt Peak National Observatory (KPNO) 4-m Mayall telescope.
The Mosaic Imager consists of 8 2k$\times$4k SITe CCDs and gives
a field of view of 36\arcmin.
Because our galaxies are small on the sky,
they were centered on one of the CCDs although the entire mosaic
of CCDs was reduced together.
Dome flats and twilight sky flats were used to determine the pixel-to-pixel variations
and scaled to remove the pupil ghost.
The images were corrected for geometrical distortions, a background
offset was removed, and the images were stacked, with cosmic ray removal,
to produce a single final image in each filter.
The Mosaic Imager images were calibrated
using field images and Landolt (1992) standard stars observed with the Lowell
Observatory 1.1 m telescope. Because IZw 115 was not observed in $B$,
a ($B-V$)$_0$ of 0.4 was assumed to calibrate instrumental $V$ magnitudes.
Uncertainties in the calibration are given in Table \ref{tab-obs}.

We obtained optical images of DDO 53, DDO 86, and NGC 4163 with the Cassegrain Focus
Imager on the KPNO 2.1 m telescope.
The T2KA CCD was used for $V$ observations
of DDO 53 and DDO 86, and T2KB was used for observations of DDO 53 in $B$ and
of NGC 4163 in $V$ and $B$.
The 2048$\times$2048 pixel format yields a 10.4\arcmin\ field of view.
Landolt (1992) standard stars were also observed for photometric calibration,
and the rms of the calibrations are given in Table \ref{tab-obs}.
We used observations of an illuminated white spot hanging in the dome (dome flats)
to remove pixel-to-pixel variations, and dark sky flats
to remove large-scale structure across the CCD.

Our experience has been that the best method for flat-fielding varies from telescope to telescope.
Discussion of this point can be found in Massey \& Jacoby (1992).
In some cases, such as at the KPNO 4-m, bright twilight flats can be used to remove both
the pixel-to-pixel variations and the large scale gradients. But in other cases, such as
the KPNO 2.1-m, twilight flats do not work well  and blank-sky flats work best.
Therefore, for the KPNO 2.1-m data, we constructed a dark sky flat for each observing run.
We obtained multiple images
with the same exposure time and moon conditions as the galaxy observations.
``Blank sky'' targets were used, but there were still many stars in the images.
The telescope was stepped between exposures and the series of images
was co-added with an algorithm that removed the stars by comparing the images
pixel by pixel and rejecting
pixels that deviated more than 3$\sigma$ from the median, where the $\sigma$ is
determined from the noise and gain parameters of the CCD.
Because the sky flats showed no structure changes from night to night,
a single sky flat was constructed for an entire observing run.
The resulting sky flat was smoothed.
We normalized the sky flat to a mean value of one near the center,
and divided the object images by this sky flat.

In the case of the $V$-band image of DDO 53, however,
we discovered after the observing run that the ``blank sky'' was contaminated
by Milky Way nebular structure (4$^{\rm h}$ 29$^{\rm m}$ 45$^{\rm s}$,
54\arcdeg 15\arcmin 33\arcsec [J2000]). The Milky Way emission produced
diffuse filamentary structure
across the field of view and so could not be used to construct
a dark sky flat. On the other hand, the galaxy itself was small and
contained in the center of the field of view; even at low light levels the galaxy
occupied only 14\% of the pixels of the image.
Therefore, for this galaxy/filter only, we used the galaxy images themselves to construct
a dark-sky flat: we combined the images to produce a single
star-free average and fit the non-galaxy portion of the image with an order 5 Legendre function,
interpolating across the area of the image containing galaxy.
The galaxy-free region was defined by looking at the extent of the galaxy with the image
displayed to see faint noise variations in the background.

We observed the galaxies only at relatively low airmass ($\leq1.4$),
high transparency, and no moonlight.
In all cases we took a series of exposures of the galaxy,
and the telescope was stepped between exposures in order to move
the galaxy around the chip and average over
flat-fielding imperfections. We only retained images with less than 7\%
transparency differences
and with point-spread-function FWHM less than 50\% greater than
the best in the series.
The flat-fielded galaxy images were shifted to align the stars and co-added to build
signal-to-noise. We used an algorithm to
remove cosmic rays and adjusted parameters so that the photometric
properties, as determined in a simple average of the images, was preserved.

In spite of our great care in determining flat-field data, the night-sky brightness at KPNO
(or any dark site) is about 22.5 mag arcsec$^{-2}$
(see, for example, Neugent \& Massey),
and so achieving 29.5 mag arcsec$^{-2}$ requires going 7 magnitudes below the
sky.  No flat-field will be good to 1 part in 6000 (10\% photometry) at these light levels.
Fortunately, it does not have to be.  The key to measuring these low light levels is not in the flat-fielding,
but rather in the sky subtraction.  The flat-fielding helps in getting the sky relatively flat from
one side of the frame to the other, but even if there are 1\% residual gradients in the flat-fielding, this is
dealt with by fitting the sky with a 2-dimensional function, as we did.

Therefore, we edited foreground stars and background galaxies from the images and
determined a two-dimensional fit of the sky pixels in the edited image---interpolating
across the galaxy.
This procedure works because the galaxies are a small portion of the total field of
view of the image: approximately 14\% for DDO 53, 12\% for DDO 86, 10\% for DDO 133,
26\% for NGC 4163, and 4\% for IZw 115 and its detached region.
It is necessary because there is structure in the sky.
Typical variations of the background across the image were of order 0.2-0.6\%.
We used a Legendre function with the minimum
order that produced a satisfactory sky. Typical orders were 3-5.
We determined the success of a fit to the sky empirically by examining
row and column cuts through the sky-subtracted image.
We expected the background to be flat with noise around zero, and we varied the
x and y orders until we achieved that.
This procedure produced an image of the sky.
We then subtracted the sky image from the original co-added image
and from the edited image to yield sky-subtracted images with and without the
foreground/background object contamination.

A legitimate question is whether or not our fits to the sky structure are affected by the presence of the galaxy
and/or contaminating stars in the field.  That can be answered best by comparing results obtained on the
same galaxy using different telescopes/instruments with differing fields of view. This
comparison is made in \S \ref{sec-optsurfphot}.
We further gain confidence in our results in the outer disks of the dwarfs from
the fact that we see similar profiles from data obtained with different telescopes/instruments
and reduced with different types of sky flats (DDO 133 and IZw 115 imaged with the Mosaic Camera
on the KPNO 4 m and reduced with twilight flats, and the other galaxies imaged with the Cassegrain Focus
Imager on the KPNO 2.1 m telescope and reduced with dark sky flats).

\subsubsection{Surface photometry} \label{sec-optsurfphot}

Removal of background galaxies and foreground stars is essential
for meaningful azimuthally-averaged surface photometry.
We attacked this problem in two ways for each image.
First, we used the edited sky-subtracted images, with some additional masking of small areas in close to the galaxy.
A problem with this approach is in distinguishing, without radial velocity information,
stellar clusters in the galaxy from foreground/background object contamination.
We used the various passband images of the galaxy at our disposal to make the best guess
as to whether a discrete object in the image
was a part of the galaxy or foreground/background contamination.
The area surrounding the galaxy is a good clue to the type of contamination we
should expect across the galaxy.
Nevertheless, subjective judgement was involved in this step.

The second method we used to account for foreground/background contamination
was to use the unedited sky-subtracted image
and determined an average ``sky'' value around the galaxy that included the
foreground/background component.
This statistically corrects
for interlopers, and assumes that the fields around the galaxy represent the contamination
across the galaxy itself.
The problem with this method is that a given annulus in the surface photometry might not contain
an exactly representative amount of light from extraneous objects, and a single
foreground star or background object
in the faint, outer parts of the galaxies can severely affect the surface photometry.
For this reason, we prefer the surface photometry obtained with the first method of editing contamination from
the images, but we applied both methods. In fact
the two methods produced similar results. The comparison is shown in Figures \ref{fig-comparemethods}
and \ref{fig-compareuvmethods} for the $V$-band and {\it GALEX} NUV photometry, respectively.
The surface photometry is cut off where there is a large jump in the uncertainty, usually around 1 magnitude arcsec$^{-2}$.

\begin{figure}
\epsscale{0.9}
\includegraphics[angle=0,width=1.0\textwidth]{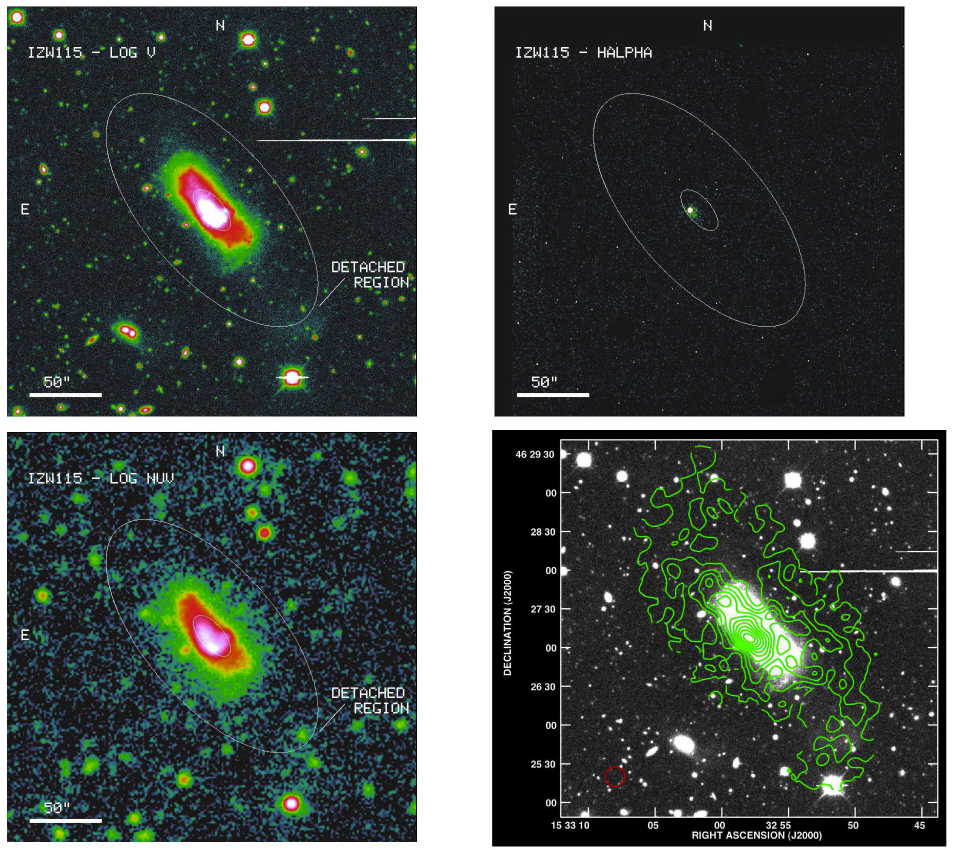}
\caption{
As in Figure 1, for IZw 115.
The H$\alpha$ image is not shown as the logarithm.
Contours of integrated \HI\ emission are
1.3 to 27.2 in steps of 2.59$\times10^{20}$ cm$^{-2}$.
The region of stars that is detached from the main body of IZw 115 is marked
in the $V$ and NUV images. The southern \protect\HI\ tail in the color composite image
extends over the detached region.
\label{fig-izw115}}
\end{figure}

\begin{figure}
\epsscale{0.9}
\includegraphics[angle=0,width=0.9\textwidth]{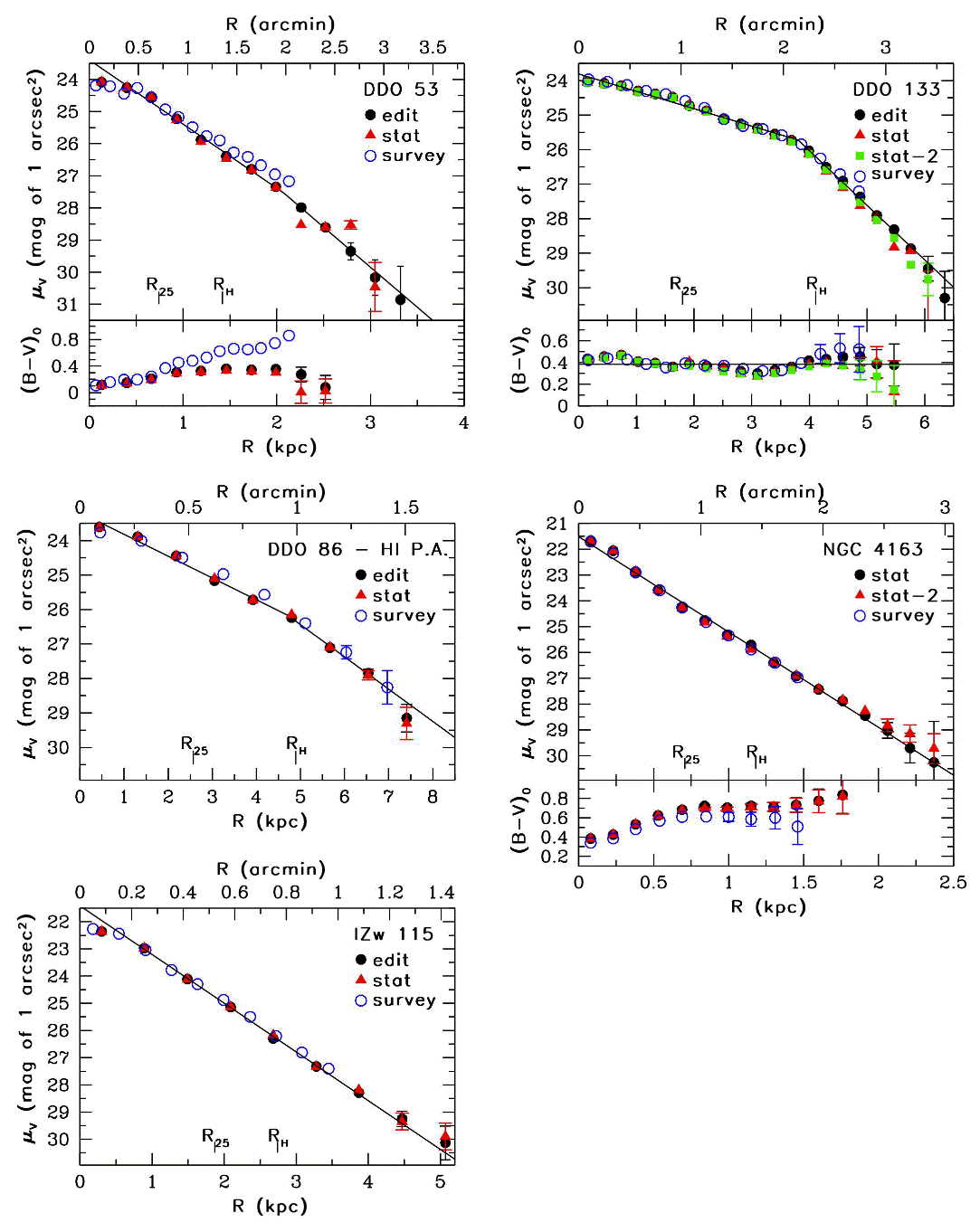}
\caption{Deep $V$, and in some cases $B-V$, surface photometry with different methods of
determining the background/foreground object contamination:
``edit" refers to editing and masking individual unwanted objects,
``stat" refers to using the surrounding field to determine a statistical correction,
and ``stat-2" refers to the statistical method carried out by a different co-author.
We also compare the deep surface photometry to
the shallower survey surface photometry from Hunter \& Elmegreen (2006).
Uncertainties are only shown where they are larger than the symbols.
$R_{25}$ and $R_H$ mark the radius at the $B$-band 25 mag arcsec$^{-2}$
isophote and the Holmberg radius ($B=$26.66 mag arcsec$^{-2}$), respectively.
\label{fig-comparemethods}}
\end{figure}

\begin{figure}
\epsscale{0.8}
\includegraphics[angle=0,width=0.8\textwidth]{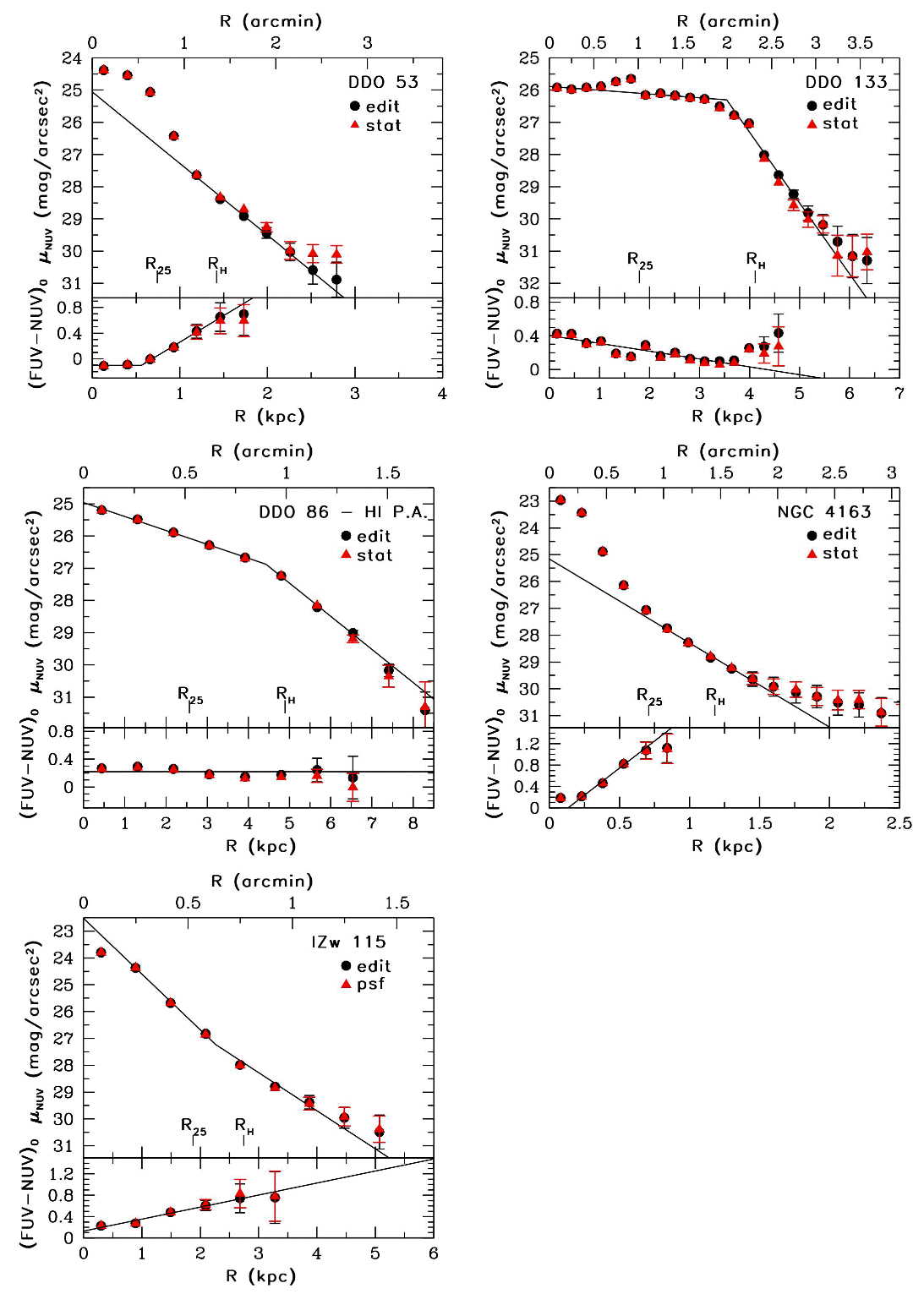}
\caption{Deep {\it GALEX} NUV and $FUV-NUV$ surface photometry with different methods of
determining the background/foreground contamination:
``edit" refers to editing and masking unwanted objects,
``stat" refers to using the surrounding field to determine a statistical correction,
and ``psf" refers to point spread function (PSF) fitting and subtraction.
Uncertainties are only shown where they are larger than the symbols.
$R_{25}$ and $R_H$ mark the radius at the $B$-band 25 mag arcsec$^{-2}$
isophote and the Holmberg radius ($B=$26.66 mag arcsec$^{-2}$), respectively.
\label{fig-compareuvmethods}}
\end{figure}

Because dIm galaxies are lumpy, we determined the center of the
galaxy, position angle (P.A.), and ellipticity from an outer isophote
in a contour plot of a smoothed version of the $V$ image. Then
we integrated the photometry in fixed ellipses that increase in semi-major axis length.
The step size is given in Table \ref{tab-obs}, and
was chosen to be $\geq$150 pc and, as seen by eye, to cover a sufficient area of the galaxy to average over
the most severe local fluctuations.
We geometrically matched the continuum-subtracted
\ha\ images to the $V$-band images and applied
the same ellipse integrations.
The outer most ellipse is shown superposed on the
$V$, NUV, and \ha\ images in Figures \ref{fig-d53} through \ref{fig-izw115}.

We corrected the photometry for reddening using the
foreground reddening, given in Table \ref{tab-sample},
plus 0.05 mag in E($B-V$) for internal reddening in $B$ and $V$ and 0.1 mag
in E($B-V$) for internal reddening in \ha.
This  level of internal reddening is consistent with measurements
of the Balmer decrement in \HII\ regions in a sample of 39 dIm galaxies (Hunter \& Hoffman 1999).
There, the average reddening in \HII\ regions is 0.1, and we have taken half this to represent
the stars outside of \HII\  regions.
We used the Cardelli \et\ (1989) extinction curve.

We used a single extinction correction for all radii within each galaxy.
This is justified by the lack of metallicity gradients in dwarfs (see, for example, Croxall \et\ 2009)
and the correlation
between metallicity and attenuation seen in spirals (Boissier \et\ 2007).
On the other hand, little is known about the outer disks of dwarfs, beyond
where \ha\ emission is no longer detected, although there are some hints of dust emission even there
(see, for example, Hinz \et\ 2006).
Boissier \et, based on {\it IRAS} observations, suggest that there is no attenuation
in the outer disks of two dwarfs (IC1613, WLM), but uncertainties are large due to low {\it IRAS}
fluxes.

There is a region of stellar emission near, but detached from, IZw 115 that we did not
include in the galactic photometry. It is located at
15$^h$32$^m$51.0$^s$ 46\arcdeg25\arcmin51\arcsec\ (2000), 101.5\arcsec\
($=$6.0 kpc) from the center of IZw 115, southwest along the galaxy's
major axis, and is marked in images in Figure \ref{fig-izw115}.
In a circular aperture of radius 14.85\arcsec, the region
has a $V$ magnitude of 20.39$\pm$0.04 and a surface brightness
$\mu_V$ of 26.9 mag of one arcsec$^2$. If it is at the distance
of IZw 115, it has an $M_V$ of $-10$ and a diameter of several kpc.
Thus, it could be a very small dwarf companion.

DDO 133 is clearly barred, and the outer isophote that was used
to determine the galaxy morphology is located beyond the bar.
The bar has a P.A. that is $-13$\arcdeg, an 11\arcdeg\
twist from the outer isophote. The center of the bar is
located at 12$^h$32$^m$54.5$^s$ 31\arcdeg32\arcmin27\arcsec,
18.8\arcsec\ ($=$560 pc) from the center of the outer isophote.
The minor-to-major axis ratio $b/a$ of the bar is 0.44, compared to
0.69 for the outer isophote, so the outer isophote is rounder.
The major axis of the bar is 133\arcsec\ ($=$3.9 kpc) long.
We experimented with changing the centering and P.A. of the photometry ellipses,
but found that there was no significant change in the photometry
for a reasonable alternate choice of the center (10\arcsec\ offset
to the southwest) and P.A. (5\arcdeg\ counter-clockwise).

DDO 86 presents a special challenge to determining the position angle
of the major axis.
At an outer isophote ($\mu_V\sim27.2$ mag arcsec$^{-2}$),
the P.A. of the major axis is 85\arcdeg\ and the
minor-to-major axis ratio $b/a$ is 0.86. This is close to
the galaxy morphology determined in the survey data:
P.A.$\sim$72\arcdeg\ and $b/a\sim$0.82 (Hunter \& Elmegreen 2006).
However, in the deeper imaging data that we have here, we
see that there is a faint extension of starlight to the north.
At an isophote level that includes this extension
($\mu_V\sim28.4$ mag arcsec$^{-2}$), the P.A. of the galaxy rotates
clockwise to 32.5\arcdeg, although the shape remains the same ($b/a\sim0.87$).
An inner isophote ($\mu_V\sim25.6$ mag arcsec$^{-2}$)
has a major axis that is rotated counter-clockwise from both of these
(113.5\arcdeg) and the galaxy is more elongated ($b/a\sim0.65$).
Finally, the kinematical major axis determined from the \HI\ velocity field has a P.A. of $-46$\arcdeg.
It is possible that this sequence of optical P.A.s---rotating
clockwise from higher surface brightness to fainter---and change
in shape---becoming rounder in the middle and outer parts---represents
an offset bar; the center of the inner isophote is
located 360 pc to the southwest of the center of the extreme outer isophote.
For the surface photometry we have chosen to use the P.A. from our \HI\ kinematic analysis.

Since we have surface photometry from the survey images of each
of these galaxies, we compare the deep imaging photometry with that
from the survey data in Figure \ref{fig-comparemethods}.
We see that the agreement is quite good in most cases. In addition, the breaks
in the surface brightness profiles seen in the survey data of
DDO 86 and DDO 133 are there at the same radius and surface
brightness level in the deep data.
And, the azimuthally-averaged $B-V$ colors in DDO 133
match closely the survey data in the region of overlap.
The exception to the good agreement is DDO 53.
Beginning at a radius of 1.2\arcmin\ (1.25 kpc) and a $\mu_V$ of 25.8
mag arcsec$^{-2}$ the new surface photometry is systematically
fainter. The differences are greater than the uncertainties
in the photometry. The survey $B-V$ color is also systematically
redder than the deep photometry beginning at a radius of
0.77\arcmin\ (0.81 kpc; $\mu_V\sim24.9$ mag arcsec$^{-2}$).
In the survey data, there is a gradient in $B-V$, with the
galaxy becoming redder with radius. In the deep data, the
color is relatively constant. We note that the integrated
$B-V$ of DDO 53 from the deep image (0.30$\pm$0.01, not corrected
for reddening) is close to that given by de Vaucouleurs \et\ (1991;
0.35), while that from the survey data (0.48) is significantly
redder. Therefore, we suspect that the deep photometry is better than
that measured from the survey data for this galaxy in the region of overlap.

From the ellipse photometry we can derive several quantities.
First, the last ellipse yields integrated photometry and that is given in Table \ref{tab-phot}.
Second, fits to the surface brightness profiles yield disk scale lengths $R_D$
and central surface brightnesses $\mu_{V,0}$.
These are listed in Table \ref{tab-disk}.
Four of the 5 galaxies---DDO 53, DDO 86, DDO 133, and NGC 4163---exhibit double exponentials.
The profile in the outer disk falls off more steeply than that in the inner disk in DDO 53, DDO 86,
and DDO 133. In NGC 4163 the outer profile is shallower than the inner.
DDO 86 and DDO 133 were already known to have double exponentials from the survey data,
but the survey data of DDO 53 did not go out far enough to give a hint of the break in the profile.
The break in NGC 4163 is more subtle than in the other three galaxies.
Nevertheless, the scale lengths are different---$0.26\pm0.02$ kpc versus $0.31\pm0.004$ kpc,
and the central surface brightnesses implied by each differ by 0.54 magnitude.
For those four galaxies with breaks, the $R_D$ and $\mu_{V,0}$ are given for both the inner and outer parts of the profiles,
along with the break radius $R_{Br}$ at which the slope changes.
In addition DDO 53 and IZw 115 show breaks at small radii, a little over one disk-scale length
that coincide with concentrations of star formation.
These are marked in the surface brightness plots, but not included in Table \ref{tab-disk}.
The surface photometry and color profiles are shown in Figures \ref{fig-d53sb} through \ref{fig-izw115sb}.

\subsection{Ultraviolet images}

We obtained deep images of our sample galaxies with the
{\it Galaxy Evolution Explorer} satellite ({\it GALEX}; Martin \et\ 2005)
to trace and characterize star formation
into the outer disks of dwarf galaxies where \ha\ may not be an effective tracer of recent star
formation (see, for example, Thilker \et\ 2005, 2007a; Boissier \et\ 2007).
The UV---ultraviolet light coming directly from OB stars---can easily
trace young stars in the outer galaxy, and
the UV has the advantage in a patchy star-forming environment of integrating over a
longer timescale than does \ha\ (10 Myrs).

The galaxies are listed in Table \ref{tab-galex} along with
the exposure times and tile name of the {\it GALEX} images.
{\it GALEX} imaged simultaneously in two channels: FUV with a
bandpass of 1350--1750 \AA, an effective wavelength of 1516 \AA, and
a resolution of 4.0\arcsec\ and NUV with a bandpass
of 1750--2800 \AA, an effective wavelength of 2267 \AA,
and a resolution of 5.6\arcsec. The images were processed through
the {\it GALEX} GR4/5 pipeline and were retrieved as final intensity maps with a
1.5\arcsec\ pixel scale.
The {\it GALEX} field of view is a circle with 1.2\arcdeg\ diameter, and we have extracted
a portion around our target galaxies.

We corrected the UV photometry for extinction using the same total reddening E($B-V$)$_t$ as
was used for correcting the optical.
We combined the E($B-V$)$_t$ with the extinction law
of Cardelli \et\ (1989) to produce the extinction
$A_{FUV}=8.24{\rm E}(B-V)_t$, and,
interpolating over the 2175 \AA\ bump, $A_{NUV}=7.39{\rm E}(B-V)_t$.
However, the FUV extinction law is known to vary from galaxy to galaxy, and
the NUV filter straddles the 2175 \AA\ extinction feature, which also varies not only from galaxy
to galaxy but from place to place within galaxies (for example, Gordon \et\ 2003).
Star formation activity appears to play as important a role, perhaps even more important,
than metallicity in determining the UV extinction curve and the strength of the 2175 \AA\ bump.
In the metal-poor SMC, for example, Gordon \& Clayton (1998) have shown that lines of sight through
the actively star forming bar have UV extinction curves that are linear with 1/$\lambda$, rising more
steeply into the FUV than in the Milky Way, and no 2175 \AA\
bump is present, while a sight line through the quiescent wing has an extinction curve that is
similar to that in the Milky Way with a shallower slope into the FUV and a 2175 \AA\ bump.
In our dwarf galaxies, the level of star formation activity is highly variable from galaxy to galaxy
and from place to place within a galaxy, potentially resulting in a variable FUV extinction law.
Wyder \et\ (2007) have dealt with this for a large sample of galaxies by
adopting the Cardelli \et\ extinction law for the Milky Way
and determining the best average reddening
from convolving theoretical galaxy spectral energy distributions (SEDs)
with the {\it GALEX} filter transmission curves.
In Hunter, Elmegreen, \& Ludka (2010),
we examined the results of convolving different SMC, LMC, and Milky Way
extinction curves (Cardelli \et 1989, Gordon  \et\ 2003)
with constant star formation rate galaxy SEDs constructed from the Bruzual \& Charlot
(2003) stellar population library and with the {\it GALEX} filter transmission curves.
The $A_\lambda/{\rm E}(B-V)$ varies by 1.5 magnitudes
depending on the extinction curve.
Considering the variety of environments represented by our sample and the fact that
extinction is minimal except for a few galaxies with higher than average foreground extinction,
we decided to adopt the Wyder \et\ extinctions:
$A_{FUV}=8.24{\rm E}(B-V)_t$ and $A_{NUV}=8.2{\rm E}(B-V)_t$.
For most of our galaxies, the variations represented by alternate extinction laws result
in differences of order 0.05 mag.

Surface photometry of the NUV and FUV images proceeded in the same manner as for the
optical images. We edited foreground/background objects from the images, fit the cleared area
around the galaxy, and subtracted this from the edited and from the unedited images.
For the unedited sky-subtracted image, we determined the average background light from
contaminating objects and subtracted this from the image.
We then geometrically transformed the UV images to match the $V$-band scale and orientation
and measured surface photometry on the edited and unedited images using the
same ellipse parameters as for the $V$-band images.
The two methods of accounting for background/foreground contamination are shown in
Figure \ref{fig-compareuvmethods}. The results of the two methods agree well.
Integrated photometry are given in Table \ref{tab-phot}.
The surface photometry and color profiles are shown in Figures \ref{fig-d53sb} through \ref{fig-izw115sb}.

Since we are looking at faint light levels in outer disks, we need to consider whether the UV light
we see truly comes from young stars or whether there might be other sources of light that dominate out there.
One such potential source is white dwarf stars.
DDO 133, as a test case, has an $FUV-NUV$ color of order 0.1 mag in the outer disk, not counting the colors with
relatively high uncertainties.
A white dwarf of that color has an effective temperature of order 16,000 K.
At a distance of 10 pc, a star with $\log g=6$ and a radius $R=R$\solar\ would have an absolute NUV AB magnitude of
$M_{NUV}=1.84$ (L.\ Bianchi, private communication).
Assuming a mass for the white dwarf of 0.7 $M$\solar, the star should have a radius of 0.1$R$\solar.
Thus, the apparent NUV magnitude would be 35.7.
To reproduce a $\mu_{NUV}$ of 28 mag arcsec$^{-2}$ at  a radius of 2$R_D$ in DDO 133,
we would need about 1.4 white dwarfs pc$^{-2}$.
That is 2 times higher than the density of white dwarfs in the solar neighborhood.
For white dwarfs with smaller radii or higher $\log g$, the number of white dwarfs that are needed goes up,
so this is probably a lower limit.
However, the solar neighborhood should have a much higher stellar density than the outer disk of a dwarf.
In the solar neighborhood $\mu_V$ is 22.7$\pm$0.2 (Ishida \& Mikami 1982),
and that is a factor of 30 higher than the surface brightness in DDO 133 at the radius we are considering ($\mu_V=26.5$).
So, the stellar density of white dwarfs needed to account for the UV light in DDO 133 fails by a factor of
about 60.
Even 3 magnitudes fainter---at the edge of our detection, the numbers needed are 4 times too high
and so it is unlikely that white dwarfs alone could produce the UV starlight that we see.
We also considered horizontal branch stars since these dominate the UV light in,
for example, M32 (Brown \et\ 2000). However, the $FUV-NUV$ color of these stars are of order 2 and redder.
This color is far redder than what we see in dwarfs.

\subsection{\HI}

Three of the galaxies in our sample---DDO 53, DDO 133, NGC 4163---are part
of the LITTLE THINGS
({\it LITTLE: Local Irregulars That
Trace Luminosity Extremes; THINGS: The HI Nearby Galaxy Survey})
project, a multi-wavelength survey of dwarf galaxies (LITTLE THINGS: Hunter \et, in preparation;
THINGS: Walter \et\ 2008).
LITTLE THINGS has \HI\ observations at additional VLA arrays, including B array,
that provide higher spatial resolution. They will provide much better maps
for analyzing the relationship between the gas and star formation.
Here we restrict ourselves to the data obtained for this project and
examine general correlations between the stars and the gas.

\subsubsection{Calibration and Mapping}

We obtained Very Large Array (VLA\footnote{
The VLA is a facility of the National Radio Astronomy Observatory (NRAO).
The National Radio Astronomy Observatory is a facility of the National Science Foundation operated under cooperative agreement by Associated Universities, Inc.})
observations of \HI-line emission in our galaxies.
The dates and configurations of the L-band observations are given in Table \ref{tab-vla}.
The total bandwidth was 1.56 MHz, and 2 IFs were used.
The data were on-line Hanning smoothed, and the channel separation was 12.2 kHz
(2.6 \kms).
The data from 2006 and 2008 were obtained when the VLA was in transition to the EVLA.
Unfortunately, in converting the signal from the EVLA antennas, power was
aliased into the lower 500 kHz of the bandwidth. The consequence of this
is that all EVLA-EVLA baselines had to be flagged and not used
(see Ficut-Vicas et al., in preparation, for a discussion of how to deal with this problem).

We subtracted the continuum emission in the {\it uv}-plane using
line-free channels on either end of the spectrum and, in the case
of DDO 53, combined the data from the two arrays.
To make maps, we employed a routine in NRAO's
Astronomical Image Processing System (AIPS)
that enables one to choose a sample weighting that is
in between the standard ``natural'' weighting, which gives the
highest signal-to-noise but at the expense of long wings in the
beam, and ``uniform'' weighting, which gives the best resolution
but at the cost of signal-to-noise and the presence of negative
sidelobes. We chose a sample weighting that gives a formal increase
of only 5\% in noise yet with a significant improvement in beam
profile over what would have been achieved with
``natural'' weighting.  The resulting synthesized beam FWHM
are given in Table \ref{tab-vla}.
We deconvolved the
maps until there were roughly comparable numbers of positive and
negative components.

To remove the portions of each channel map without emission that
are only contributing noise to the map, we used
the maps smoothed to twice the beamsize for a conditional transfer of
data in the unsmoothed maps. The channel maps were blanked wherever
the flux in the smoothed maps fell below 2.5$\sigma$.
Flux-weighted moment maps were made from the resulting data cube.
Integrated emission---moment zero---maps are shown in Figures \ref{fig-d53vf} through \ref{fig-izw115vf},
along with velocity field (moment one) and velocity dispersion (moment two) maps.
Integrated \HI\ contours superposed on the $V$-band images are shown in
Figures \ref{fig-d53} through \ref{fig-izw115}.

\subsubsection{Velocity fields}
\label{sec-vf}

Position velocity diagrams (panel e), velocity fields (panels b and d), and velocity dispersions
(panel c) are shown in Figures \ref{fig-d53vf} to \ref{fig-izw115vf}.
Panel (b) shows the standard intensity-weighted mean velocity field, and panel (d) shows the velocity field that
results from fitting profiles with a Gauss-Hermite $h_3$ polynomial.

Because of the possible presence of multiple velocity components, we have
deconvolved ordered from non-ordered motions using the method
described in Oh \et\ (2008). First, an approximate rotation curve is derived from the
major axis position-velocity diagram and this is used to create an artificial velocity field.
Then in an automated method, we fit single Gaussians to the major velocity component in each profile
across the galaxy. The velocities of these components are compared to the artificial velocity field.
If they deviate more than a prescribed amount, the program examines the possibility that the bulk motion at
that point is represented by a secondary velocity component.
From this, we extract the bulk velocity field that then
becomes the input velocity field for the next iteration, which is now fit using the ``tilted ring model'' rather than
the major axis position-velocity diagram.
This loop continues until the mean difference between successive velocity fields is less than
three times the channel width.
This gives a final bulk velocity field.
Parameters derived from this process (center position, P.A., inclinations)
are summarized in Table \ref{tab-himaps}.
Following Bureau \& Carignan (2002) we corrected the rotation velocities for asymmetric drift, and the
final rotation curves are shown in the graphs in Figures \ref{fig-d53sb} through
\ref{fig-izw115sb}.

The velocity fields of DDO 86 and DDO 133 exhibit well-behaved ordered rotation.
In IZw 115 the velocity field is less well-behaved, but still ordered motion is present.
The velocity fields in DDO 53 and NC 4163, on the other hand, are quite peculiar.
Here we discuss each galaxy individually:

{\it DDO 53}.
DDO 53 is a part of the THINGS study of \HI\ in nearby galaxies (Walter 2008) and has
been analyzed by Oh \et\ (in preparation).
The kinematical axis of DDO 53 is at a position angle of 155\arcdeg, while the optical
major axis is at an angle of 81\arcdeg, a difference of 74\arcdeg. Furthermore, the
fit to the \HI\ velocity field yields a variable inclination angle. The kinematic center
and the optical center are 32\arcsec\ apart, with the kinematic center being NE of the optical
center. The fit to the velocity field is quite poor, and the rotation curve is only meaningful out to
a radius of 0.9 kpc.

{\it DDO 86}.
The bulk velocity field of DDO 86 is well determined with clean fits to the position angle of the kinematic axis,
the center position, and the inclination of the \HI\ disk. The optical, however, is more problematic, as discussed
in \S \ref{sec-optsurfphot}: The $V$-band major axis rotates
clockwise from higher surface brightness to fainter. This may indicate the presence of an off-center bar.
So, for the optical and UV surface photometry we used the P.A.\  derived from the \HI\ kinematics.
The optical center determined from outer isophotes agrees within 14\arcsec\ with the \HI\ center,
with the \HI\ center being NW of the optical.

{\it DDO 133}.
The bulk velocity field of DDO 133 is well-behaved. The kinematic P.A., inclination,
and center agree very well with those determined for the optical.

{\it NGC 4163}.
Although rotation is seen in the gas, the velocity field is so peculiar that parameters for the tilted
ring model---center, P.A., and inclination--were taken from the optical.

{\it IZw 115}.
IZw 115 shows rotation, although the velocity field is rather peculiar. The position angle of the kinematic
axis was found to vary between 50\arcdeg\ and 78\arcdeg, while the optical major axis is at an angle
of 41\arcdeg. On the other hand, the inclinations agree within 7\arcdeg, and the centers, within 4\arcsec.
The detached region seen in the optical and NUV (Figure \ref{fig-izw115}) shows a hint of rotating
independently of the main body of IZw 115. Panel (d) in Figure \ref{fig-izw115vf},
which shows the velocity field resulting from fitting profiles with a Gauss-Hermite $h_3$ polynomial,
can be interpreted as showing overall rotation with an axis at a
position angle of roughly 90\arcdeg, clearly distinct from IZw 115. This could indicate that the detached region
is a small companion galaxy that is interacting with IZw 115.

\subsubsection{Surface density profiles}

The azimuthally-averaged \HI\ surface
densities $\Sigma_{HI}$ and velocity dispersions $\sigma_{vel}$ of the rotational component were determined using
the center, P.A., and inclination angle given in Table \ref{tab-himaps}.
The $\Sigma_{HI}$ was corrected for He
to produce
the gas surface density $\Sigma_{gas}$.
The gas surface density profiles and average velocity dispersions
are shown as a function of radius in Figures \ref{fig-d53sb} through \ref{fig-izw115sb}.

\section{Stars in the outer disks}
\label{sec-stars}

\subsection{Where do dIm stellar disks end?}

We have traced starlight in the $V$-band to ultra-low surface brightness levels.
Yet, the stellar surface brightnesses in the outer disks of our 5 galaxies
continue to decline exponentially as far as we can measure it.
That is, we run out of signal-to-noise before we convincingly run out of galactic starlight.
We have not found the edges of the stellar disks.
This is consistent with results from stellar photometry that have traced
the stellar disks in the LMC (Saha \et\ 2010), M33 (Barker \et\  2011), and M31 (Richardson \et\ 2008)
to even more extreme radii.

\subsection{Breaks in the Surface Brightness Profiles}

The surface photometry of DDO 133 shows a change in the slope at 1.8$R_D$.
Beyond there, $\mu_{V}$ drops more steeply.
Although less extreme, there are noticeable changes in the slopes in the outer disks of DDO 53, DDO 86,
and NGC 4163 as well.
And, in DDO 86 and DDO 133 the breaks are also apparent in NUV surface
brightness profiles.
The break in NGC 4163 is coincident with the outer limit
to a central bump in NUV and \ha\ emission. And in DDO 86 and DDO 133,
the outer disk breaks occur where \ha\ ends or declines sharply.
In addition to the breaks at several disk scale lengths, there are breaks
or variations in the surface brightness profiles at smaller radii.
We see these in DDO 53 and IZw 115. The inner bump in the $V$-band
surface photometry in DDO 53 is mimicked in NUV and \ha\ and corresponds,
more or less, to a change in FUV$-$NUV and $B-V$ colors.
The flattening out of the $V$-band surface brightness profile in the central 1.5$R_D$
in IZw 115 is accompanied by a flattening in the NUV profile, and the location of
a single bright \HII\ region.
The only break that does not coincide with a change in \ha\ is the outer break in DDO 53.
Thus, most of these breaks appear to be connected with changes in the star formation activity.

Double-exponential surface brightness profiles are also seen in about
1/4 of our survey of dIm and Sm galaxies (Hunter \& Elmegreen 2006) and
in the outer parts of many spirals (van der Kruit \& Shostak 1982;
Shostak \& Van der Kruit 1984; de Grijs \et\ 2001; Kregel \et\ 2002;
Pohlen \et\ 2002; Kregel \& van der Kruit 2004), including one low
luminosity spiral (Simon \et\ 2003). It is now also seen in disks at
high redshifts ($0.6<z<1.0$; P\'erez 2004). Traditionally, these
surface brightness breaks have been called ``truncations,'' but this
implies that the break signals the edge of the stellar disk. In dwarf
galaxies, this does not seem to be the case; the break is a rather
abrupt change rather than an end. But what changes at that radius?
Elmegreen \& Hunter (2006) have suggested that the primary driver of
star formation changes at the break, from gravitational instabilities
in the interior to turbulence compression and supernovae in the outer
disk. The nature of the breaks in dwarfs is explored further with our
much larger survey data by Herrmann \et\ (in preparation).

\subsection{What is the star formation history in the outer disks?}

What does our observational limit of 29.5 mag arcsec$^{-2}$ in $\mu_V$ mean?
Let us take DDO 133 at a distance of 6.1 Mpc as an example.
An annulus centered at $\mu_V=29.5$ mag arcsec$^{-2}$ with a width of 1 kpc would
have an area of 45617 arcsec$^2$. This gives an $M_V$ of $-11.1$ and
an $L_V$ of $2.3\times10^6 L$\solar.
We measure a $B-V$ of order 0.3-0.4 in DDO 133.
A $B-V$ of 0.35 gives a mass to light ratio $M/L_V$ of 1.07,
under the assumption of a constant star formation rate and Salpeter (1955) stellar initial
mass function (Bell \& de Jong 2001).
Using this stellar mass to light ratio, we obtain a mass
of stars in the annulus of $2.5\times10^6$ M\solar. The
stellar density is 0.06 M\solar\ pc$^{-2}$, or one Sun every 16 pc$^{2}$.

To look at ages of the stars in the outer disks,
we consider an FUV$-$NUV color of 0.1, which is the value measured
in the outer disk of DDO 133 (1-1.7$R_D$), before the uncertainties in the color rise.
At solar metallicity, an FUV$-$NUV color of 0.1 magnitude, implies a single age population
that is 300 Myrs old (Bruzual \& Charlot 2003).
But at $Z=0.004$ the age increases to 670 Myrs and at $Z=0.0004$
to 1 Gyr. At lower metallicities, stars become bluer in FUV$-$NUV and so older stars
take on the colors of that one would find for younger stars in a higher metallicity population.
Thus, at our metallicities, the stellar populations are not likely to be as young as we
might otherwise predict.

To look at the rates necessary to form the stars we find in the outer disks,
we have computed azimuthally-averaged profiles of
star formation rates (SFRs) derived from \ha, FUV, and $V$-band luminosities:
$SFR_{\rm H\alpha}$, $SFR_{\rm FUV}$, and $SFR_{\rm V}$.
We use the formulae outlined in Hunter \et\ (2010) with a Salpeter (1955) stellar
initial mass function (IMF) and assuming a constant SFR over long timescales.
For the FUV and \ha\ formulae, we use the formulae
from Kennicutt (1998) modified for the sub-solar ($Z=0.008$) metallicities appropriate for dwarf galaxies.
To determine the SFR averaged over a longer time scale, we use the $V$-band luminosity to
determine the mass in stars and assume a timescale of 12 Gyr for the formation of that stellar mass.
We use a stellar $M/L_V$ ratio from Bell \& de Jong (2001)
that is a function of the $(B-V)_0$ color.
If there is no $(B-V)_0$ measurement or
the uncertainty in the color is $>0.2$ magnitude, we assumed the last good measurement
of $B-V$. In the case of DDO 86 and IZw 115, for which we do not have deep $B$-band images,
we used the survey data. For DDO 86 $B-V$ is constant with radius in the survey photometry;
for IZw 115 the color reddens with radius and we used the survey photometry as far out as it goes.
The SFR determined from the $V$-band
is the SFR necessary to produce the original mass in stars.
We correct this by a factor of 2 for the mass recycled back into the interstellar medium
from Brinchmann \et\ (2004).

Now we consider the effects of metallicity and a further correction to
the SFR formulae based on the actual metallicities of our sample.
Two of the galaxies in our sample have oxygen abundances  $12+\log {\rm O/H}$ measured from \HII\
regions: The measured value of the oxygen abundance in DDO 53 is 7.5$\pm$0.1 according to
Hidalgo-G\'amez \et\ (2003) and $7.82\pm0.09$ according to Croxall \et\ (2009). Given the coursness of
the metallicity grid of the models, both measurements give the same SFRs.
In DDO 133 the oxygen abundance is 8.0$\pm$0.05 (Hunter \& Hoffman 1999).
For the other galaxies, there are no measured values, so we have estimated the oxygen
abundance from $M_B$ and Richer \& McCall's (1995) correlation between $M_B$ and $12+\log {\rm O/H}$.
This yields $12+\log {\rm O/H}$ of 8.0 for DDO 86 and IZw 115 and 7.7 for NGC 4163.
Metallicities are unknown in the outer disks, but are likely to be lower than these values.
To modify our star formation formulae for these metallicities, we use the
synthesis models of STARBURST99 (Leitherer \et\ 1999), with constant star formation rates and Salpeter
stellar IMFs. Their Figure 54 gives the luminosity at 1500 \AA\ for $Z$ of 0.008, 0.004, and 0.001, and
this allows us to adjust our $Z=0.008$ formulae for FUV-based SFRs
to lower metallicities. Similarly, their Figure 78, which gives the number of photons below 912 \AA,
enables us to correct our \ha\ formula, and finally Figure 58 allows us to correct $B-V$, which
enters into the formula for $M/L_V$ and the $V$-band SFR.

For our galaxies, the SFRs at a radius where $\mu_V=29.5$ mag arcsec$^{-2}$
are given in Table \ref{tab-sfrs}. Only $SFR_{\rm FUV}$ and $SFR_{\rm V}$ are given
since \ha\ is not detected at those radii.
In DDO 133, for example, the stars continue for at least another 1.5 kpc beyond where \ha\ emission
is no longer detectable.
In DDO 133 at an annulus centered at $\mu_V=29.5$ mag arcsec$^{-2}$,
the SFR determined from the stellar mass $SFR_{\rm V}$ is 0.00001 M\solar yr$^{-1}$ kpc$^{-2}$.
That corresponds to 0.0004 M\solar yr$^{-1}$ in the 1-kpc wide annulus we are using
as an example. This corresponds to approximately 6 Orion Nebulae over an area of 40 kpc$^{2}$.
This is the level of activity that we should expect to find on average at any given moment.

The azimuthally-averaged SFR profiles are shown in Figure \ref{fig-sfrrat} as ratios of
$SFR_{\rm H\alpha}$ and $SFR_{\rm FUV}$ to $SFR_{\rm V}$.
We are able to measure SFRs from FUV and $V$-band images out to 3-7 disk scale lengths.
Again, H$\alpha$ is not detected as far out as $V$ and $FUV$ emission, and so is not
a good indicator of star formation in the outer disk (see Hunter \et\ 2010 for
a discussion of this issue, as well as Meurer \et\ 2009 and Lee \et\ 2009).
However, in the region where \ha\ and FUV emission overlap,
the two measures of SFRs track each other well except in NGC 4163.
This is obvious to the eye as parallel curves in Figure \ref{fig-sfrrat}.
In addition, within the likely systematic uncertainties, the absolute levels of the \ha\ and FUV SFRs
are the same except in the center of DDO 53, within 1.5$R_D$ of NGC 4163, and in IZw 115.
In those cases the \ha\ SFR is significantly lower than the FUV SFR, implying a higher SFR
several 100 Myrs ago compared to the most recent 10 Myrs.

These three measures of the SFR can track star formation over different
time scales: 10 Myrs, several times 100 Myrs, and over 1000 Myrs,
respectively for $SFR_{\rm H\alpha}$, $SFR_{\rm FUV}$, and $SFR_{\rm
V}$. What then do these profiles tell us about the radial variations in
the star formation histories in our five galaxies?

{\it DDO 53}.
In the central 2$R_D$ of DDO 53 the SFR over the recent 10-300 Myrs
has been higher than the average in the past by factors of 3-4, while in the outer disk the SFR is suppressed
by similar factors compared to the past average.
Recent star formation in DDO 53 has retreated to the central regions.
This is also reflected in the color profiles: FUV$-$NUV is constant in the inner disk scale length
and becomes progressively redder beyond that; $B-V$ is slightly bluer in the central regions.

{\it DDO 86}.
Star formation in DDO 86 has been approximately constant over the past Gyr at all radii with only
a slight depression of the current SFR relative to the average past rate in the center.
Similarly, the FUV$-$NUV color is nearly constant with radius.

{\it DDO 133}.
In the central 0.5$R_D$ of DDO 133 the SFR over the recent 10-300 Myrs
is depressed compared to the average in the past by factors of several, but beyond that
the SFR has been roughly constant until it drops a bit beyond 2$R_D$.
The FUV$-$NUV color is red in the center and becomes steadily bluer with radius until
it turns a bit redder again at 2$R_D$.
$B-V$, on the other hand, is nearly constant with radius to 2.5$R_D$.
This flat color profile implies that the longer timescale star formation history
has been the same throughout the entire galaxy, including the outer edges.

{\it NGC 4163}.
In the central 1$R_D$ the SFR averaged over the past several 100 Myrs is comparable to that
measured over the past 1 Gyr, but the SFR measured over the past 10 Myrs is down by
a factor of about 10. Thus, in the center of NGC 4163 star formation has recently declined
significantly. Outside of the central 1$R_D$, the recent star formation activity is depressed
relative to the past average by a factor of up to 30. It seems that only the center of this
galaxy is currently engaged in significant star formation.
In terms of the colors, FUV$-$NUV becomes steadily redder with radius beyond 1$R_D$.
$B-V$ becomes redder to 2.5$R_D$ and then is constant with radius.

{\it IZw 115}.
IZw 115 is a Blue Compact Dwarf, which are often assumed to be starburst or post-starburst
systems. In the case of IZw 115, however, the $SFR_{FUV}$ is comparable to the average past rate
in the central 2$R_D$ and is depressed by factors of a few in the outer disk.
However, the most recent star formation, as seen in \ha, is down by a factor of order 10 where
it is detected in the inner 2.5$R_D$.
Thus, star formation in the center of IZw 115 was much higher in the past several 100 Myrs than it is now.
In that sense, it is post-starburst.
The FUV$-$NUV color reflects this, becoming steadily redder with radius.

\section{Gas Morphology and Kinematics: No Normal Galaxies}
\label{sec-gas}

\subsection{DDO 86 and a Hole in the ISM}

The \HI\ distribution in DDO 86 is dominated by a central hole with a higher density gas ring around it.
The hole is not as distinct as a simple circular region of low gas density;
there are several minima and some gas ``clouds'' within the shell.
But the general hole/ring morphology is clearly present.
The hole, measured from the inner edge of the shell, has a diameter of order 5.2 kpc,
the structure, measured from the midpoint of the shell, has a diameter of order 6.9 kpc,
and the shell itself is of order 2.0 kpc thick. The mass of \HI\ in the shell,
about $4\times10^8$ M\solar, is 41\% of the total mass of the galaxy.
Many other dwarf galaxies exhibit this type of morphology
(see, for example, Dopita \et\ 1985; Puche \et\ 1992; Martin 1998; Walter \&
Brinks 1999; Parodi \& Binggeli 2003; Cannon \et\  2005; de Blok \& Walter 2006; Cannon \et\ 2011a,b).
One galaxy with a similar appearance
and similar hole/shell dimensions and relative mass is DDO 88 (Simpson \et\  2005).

Although DDO 86 has a beautifully normal bulk rotation velocity field, at almost every point
within the shell and hole there are weak non-circular motions as well. In particular there appears
consistently to be a component that is $\sim$10-25 \kms\  off the rotation speed at a given position.
This suggests that the shell is young enough to still be expanding at 10-25 \kms.
McCray and Kafatos' (1987) model of stellar blown shells give an age of 60-150 Myrs for a shell
observed with an expansion velocity of 10-25 \kms\ and a radius of 2.6 kpc.
If the ambient gas density before the shell formed was of order 0.1 cm$^{-3}$,
100-1600 stars with masses greater than 18 M\solar\
would have been required to produce the shell.
That means that a substantive burst of star formation took place in the center of DDO 86 in
the past 200 Myrs.

However, neither the FUV$-$NUV color in Figure \ref{fig-d86sb} nor the SFRs in Figure \ref{fig-sfrrat}
is consistent with this picture.
The \HI\ minima regions in the center yield an FUV$-$NUV of 0.24$\pm$0.02,
which indicates an age of order 550 Myrs for a Chabrier stellar initial mass function,
a single starburst population, and a metallicity of 0.008 (Bruzual \& Charlot 2003).
A single age stellar population at 200 Myrs of age should have an FUV$-$NUV color of order $-0.4$.
Similarly, the SFR profiles shown in Figure \ref{fig-sfrrat} are counter to expectations:
The SFR$_{FUV}$/SFR$_V$ and SFR$_{H\alpha}$/SFR$_V$ ratios should show a bump up in the center,
indicating a higher recent SFR compared to the average over a long timescale.
Instead these ratios show a slight dip in the center.
These discrepancies could be due to a combination of factors: for example, a more complex star formation history
(see, for example, Cannon \et\ 2011),
contamination of the younger stellar population by an older underlying disk, local extinction, and
UV calibration uncertainties (see Zhang \et, in preparation).

\subsection{Stellar Bars}

One of our galaxies clearly has a stellar bar and a second galaxy probably has one:
The bar in DDO 133 is obvious from a casual
look at the $V$-band image, while that in DDO 86 is suspected because
the deep $V$-band image shows a different isophote shape compared to
shallower images. Both of these galaxies have normal bulk rotation fields.
In DDO 133 we see a slight S-shape to the iso-velocity contours (Figure \ref{fig-d133vf})
that is characteristic of a barred disk. Such a feature is not obvious for
DDO 86 (Figure \ref{fig-d86vf}). Interestingly, DDO 133 also has
UV-bright clumps all the way out to 2 disk scale-lengths and large
UV-bright regions are found especially at the ends of the bar.
\HI\ profiles are double-peaked in the \HI\ minimum that corresponds to a concentration
of UV clumps on the northeastern edge of the bar.
And \HI\ profiles show broad features, probably due to multiple components,
in the \HI\ minima to the east of the bar and in the region of bright UV emission
to the south of the bar.

\subsection{Peculiar Kinematics}

Two of our galaxies are particularly kinematically peculiar. Extracting the bulk rotation from
DDO 53 and NGC 4163 was difficult and the rotation curve for DDO 53 is
not meaningful outside a radius of 0.9 kpc (but see Oh \et\ 2011).
Furthermore, the rotation speeds
that were measured are small, less than 10 \kms.
In the outer disk, the \HI\ associated with NGC 4163 sports two ``spiral arm''
like features---one on the southern edge extending to the west and the other
on the northern edge pointing to the NE (Figure \ref{fig-n4163}).
In addition, the iso-velocity contours show a strange extension to the SW
(Figure \ref{fig-n4163vf}).
\HI\ profiles show complex, multiple (3-4) components outside of the central \HI\ bright region,
especially in the \HI\ extension to the South of center and along the ridge to the West of
the central region.
In DDO 53, the region of highest velocity dispersion is found along a grand arc
on the SE edge of the \HI\ distribution, where the iso-velocity contours
are most bunched together (Figure \ref{fig-d53vf}).
Except for the southern of the two \HI\ clumps, the \HI\ profiles are everywhere
broad and complex with multiple components.
In neither galaxy is it obvious from the optical morphology that anything
unusual is going on in these galaxies other than the fact that the
current star formation is concentrated to the center in both systems.

IZw 115 potentially has an excuse at disturbed kinematics, and the bulk
velocity field, although more obvious than in DDO 53 and NGC 4163, is not clean.
The detached region
to the SW (Figure \ref{fig-izw115}) could be a tiny companion interacting
with IZw 115. The gas associated with IZw 115 extends to that region,
and the region itself seems to be rotating at a different position angle
from the main body of IZw 115.
The signal-to-noise is low, but the \HI\ profiles are everywhere complex, indicating
multiple (3-4) velocity components.
IZw 115 has one central star forming complex, as is characteristic of Blue
Compact Dwarfs.

\section{Comparison of stars, gas, and star formation}
\label{sec-sf}

\begin{figure}
\epsscale{1.0}
\includegraphics[angle=0,width=0.8\textwidth]{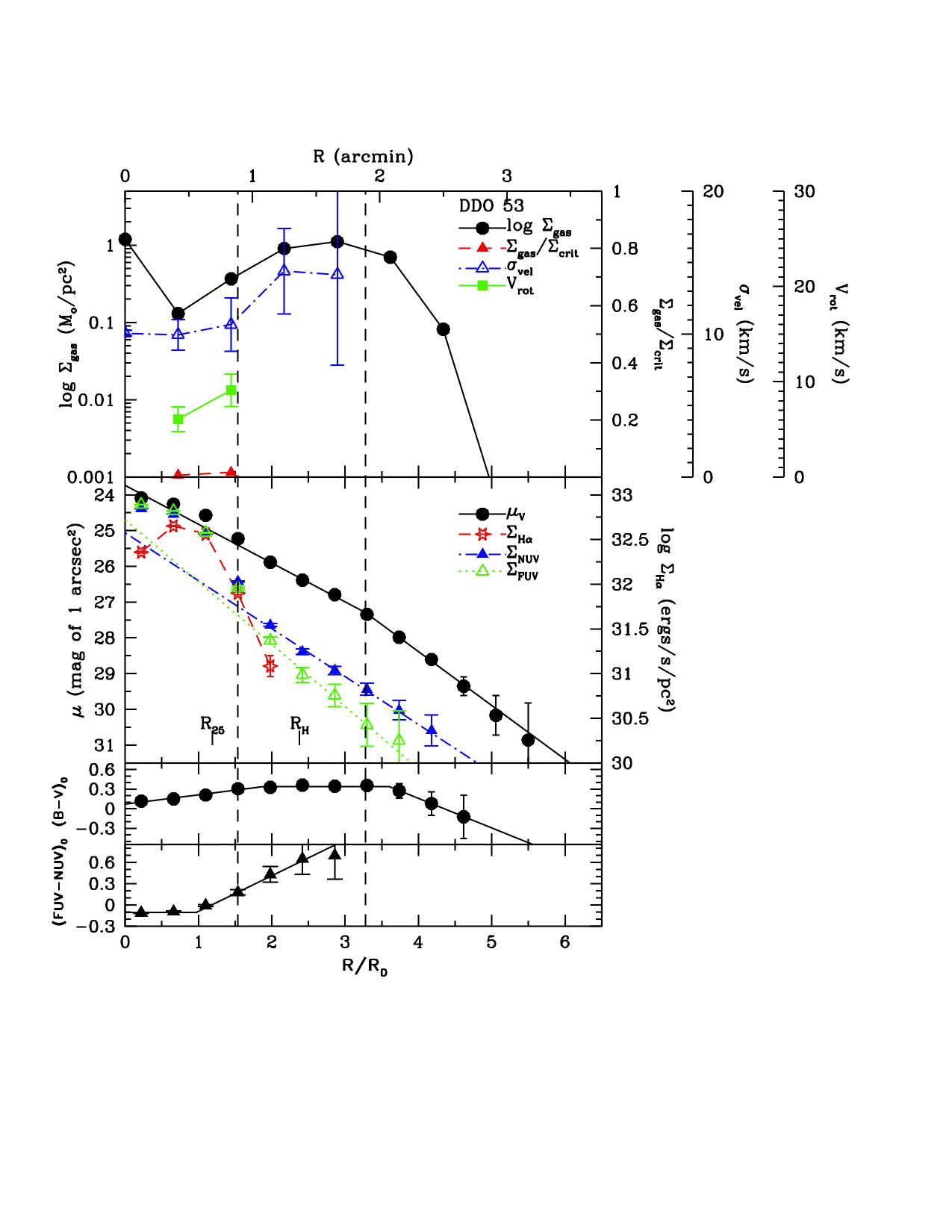}
\caption{
Azimuthally-averaged surface photometry and colors of DDO 53, corrected for reddening, plotted against radius
from the galaxy center normalized to the disk scale-length $R_D$.
{\it Top}: \protect\HI\ properties, including the gas surface density,
rotational velocity, velocity dispersion of the rotational component,
and the ratio of gas surface density to critical density.
$\Sigma_{crit}$ is the critical threshold density for large-scale
gravitational instabilities in a differentially rotating thin disk (Toomre 1964).
$\Sigma_{gas}$ includes \protect\HI\ and He.
{\it Bottom}: Optical and ultraviolet properties, including $V$, \protect\ha, and $NUV$ surface photometry.
$R_{25}$ and $R_H$ mark the radius at the $B$-band 25 mag arcsec$^{-2}$
isophote and the Holmberg radius (26.66 mag arcsec$^{-2}$), respectively.
The dashed vertical black lines mark breaks in the $V$-band profile.
\label{fig-d53sb}}
\end{figure}

\begin{figure}
\epsscale{1.0}
\includegraphics[angle=0,width=0.8\textwidth]{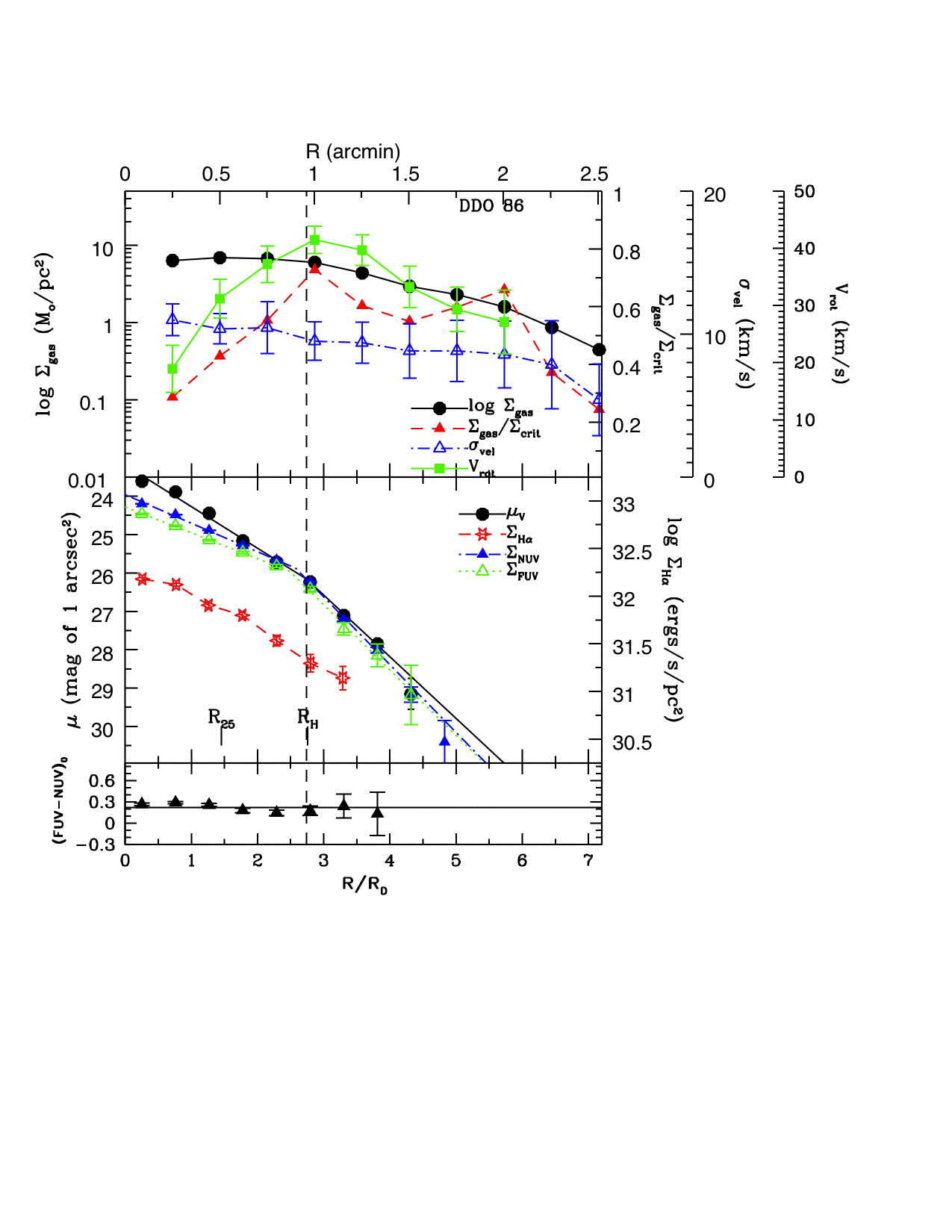}
\caption{
Azimuthally-averaged surface photometry and colors of DDO 86, corrected for reddening, plotted against radius
from the galaxy center normalized to the disk scale-length $R_D$.
{\it Top}: \protect\HI\ properties, including the gas surface density,
rotational velocity, velocity dispersion of the rotational component,
and the ratio of gas surface density to critical density.
$\Sigma_{crit}$ is the critical threshold density for large-scale
gravitational instabilities in a differentially rotating thin disk (Toomre 1964).
$\Sigma_{gas}$ includes \protect\HI\ and He.
{\it Bottom}: Optical and ultraviolet properties, including $V$, \protect\ha, and $NUV$ surface photometry.
$R_{25}$ and $R_H$ mark the radius at the $B$-band 25 mag arcsec$^{-2}$
isophote and the Holmberg radius (26.66 mag arcsec$^{-2}$), respectively.
The dashed vertical black line marks the break in the $V$-band profile.
\label{fig-d86sb}}
\end{figure}

\begin{figure}
\epsscale{1.0}
\includegraphics[angle=0,width=0.8\textwidth]{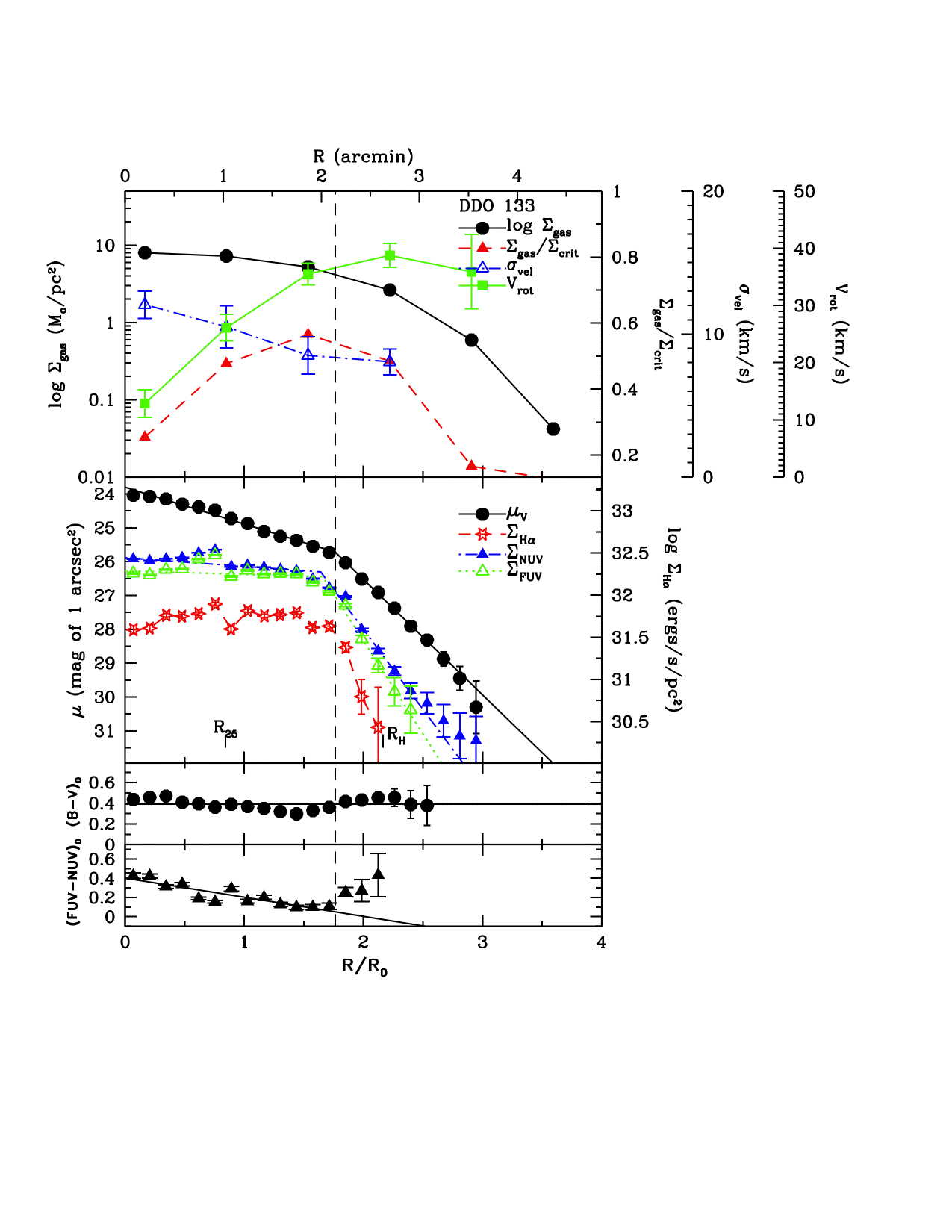}
\caption{
Azimuthally-averaged surface photometry and colors of DDO 133, corrected for reddening, plotted against radius
from the galaxy center normalized to the disk scale-length $R_D$.
{\it Top}: \protect\HI\ properties, including the gas surface density,
rotational velocity, velocity dispersion of the rotational component,
and the ratio of gas surface density to critical density.
$\Sigma_{crit}$ is the critical threshold density for large-scale
gravitational instabilities in a differentially rotating thin disk (Toomre 1964).
$\Sigma_{gas}$ includes \protect\HI\ and He.
{\it Bottom}: Optical and ultraviolet properties, including $V$, \protect\ha, and $NUV$ surface photometry.
$R_{25}$ and $R_H$ mark the radius at the $B$-band 25 mag arcsec$^{-2}$
isophote and the Holmberg radius (26.66 mag arcsec$^{-2}$), respectively.
The dashed vertical black line marks the break in the $V$-band profile.
\label{fig-d133sb}}
\end{figure}

\begin{figure}
\epsscale{1.0}
\includegraphics[angle=0,width=0.8\textwidth]{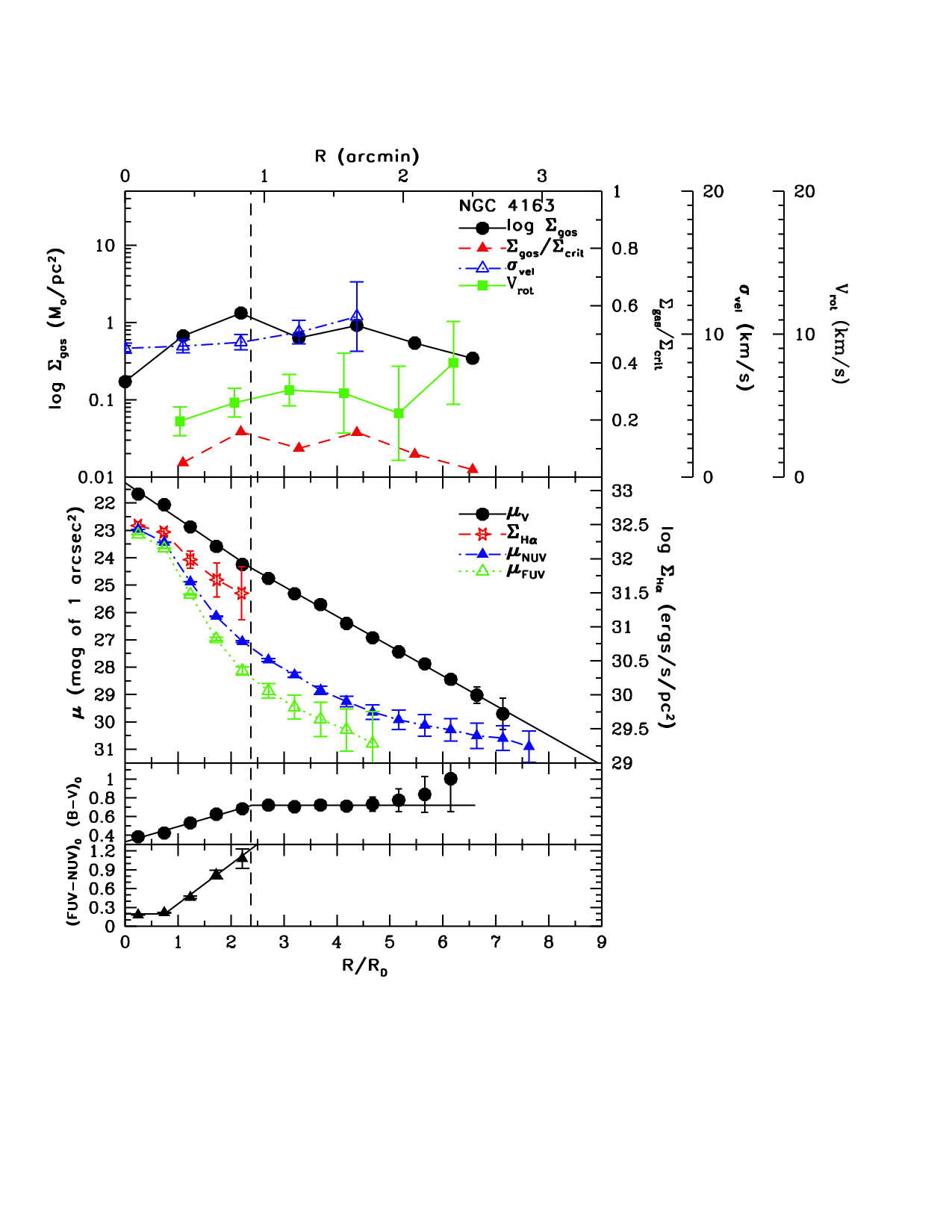}
\caption{
Azimuthally-averaged surface photometry and colors of NGC 4163, corrected for reddening, plotted against radius
from the galaxy center normalized to the disk scale-length $R_D$.
{\it Top}: \protect\HI\ properties, including the gas surface density,
rotational velocity, velocity dispersion of the rotational component,
and the ratio of gas surface density to critical density.
$\Sigma_{crit}$ is the critical threshold density for large-scale
gravitational instabilities in a differentially rotating thin disk (Toomre 1964).
$\Sigma_{gas}$ includes \protect\HI\ and He.
{\it Bottom}: Optical and ultraviolet properties, including $V$, \protect\ha, and $NUV$ surface photometry.
$R_{25}$ and $R_H$ mark the radius at the $B$-band 25 mag arcsec$^{-2}$
isophote and the Holmberg radius (26.66 mag arcsec$^{-2}$), respectively.
The dashed vertical black line marks the break in the $V$-band profile.
\label{fig-n4163sb}}
\end{figure}

\begin{figure}
\epsscale{1.0}
\includegraphics[angle=0,width=0.8\textwidth]{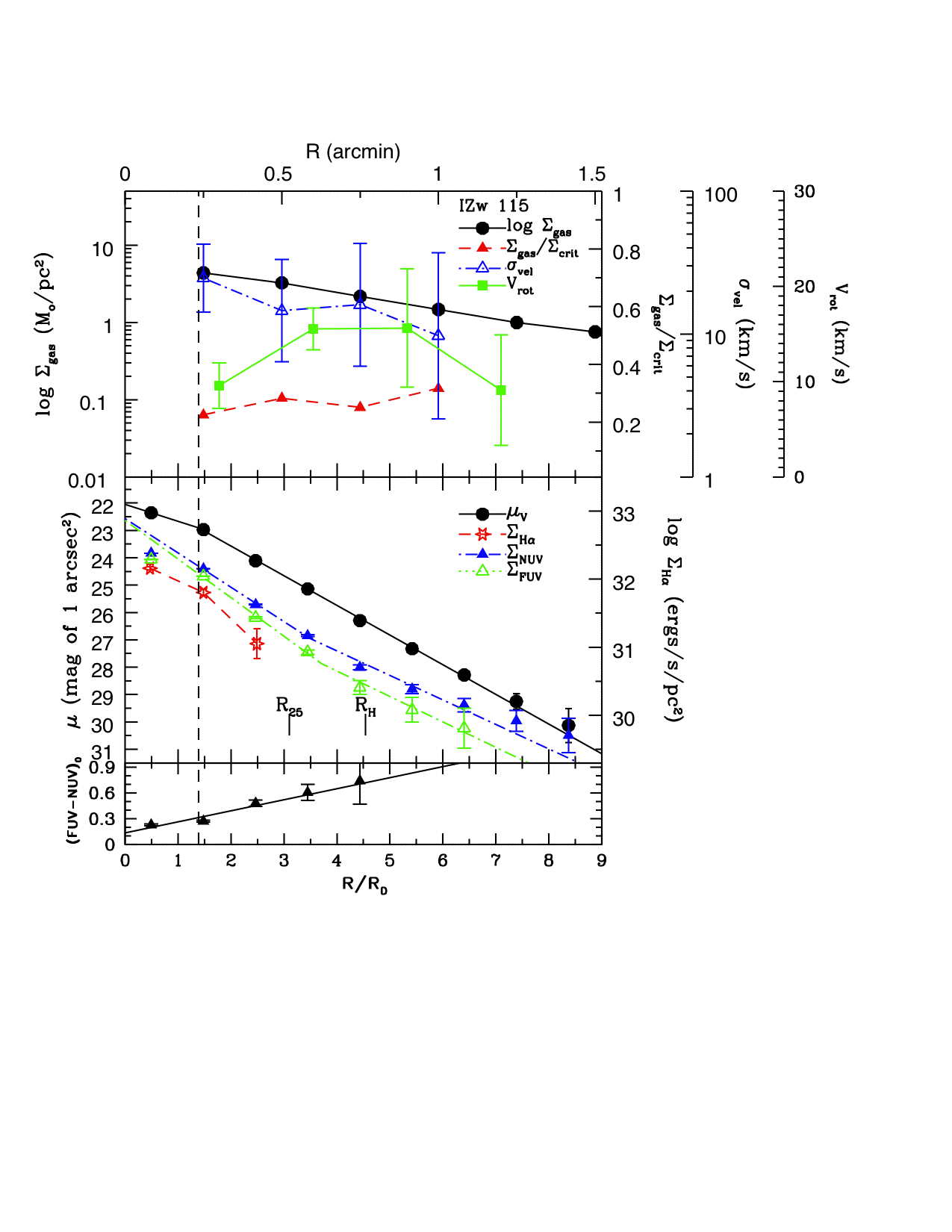}
\caption{
Azimuthally-averaged surface photometry and colors of IZw115, corrected for reddening, plotted against radius
from the galaxy center normalized to the disk scale-length $R_D$.
{\it Top}: \protect\HI\ properties, including the gas surface density,
rotational velocity, velocity dispersion of the rotational component,
and the ratio of gas surface density to critical density.
$\Sigma_{crit}$ is the critical threshold density for large-scale
gravitational instabilities in a differentially rotating thin disk (Toomre 1964).
$\Sigma_{gas}$ includes \protect\HI\ and He.
{\it Bottom}: Optical and ultraviolet properties, including $V$, \protect\ha, and $NUV$ surface photometry.
$R_{25}$ and $R_H$ mark the radius at the $B$-band 25 mag arcsec$^{-2}$
isophote and the Holmberg radius (26.66 mag arcsec$^{-2}$), respectively.
The dashed vertical black line marks the break in the $V$-band profile.
\label{fig-izw115sb}}
\end{figure}

\begin{figure*}
\includegraphics[angle=0,width=1.0\textwidth]{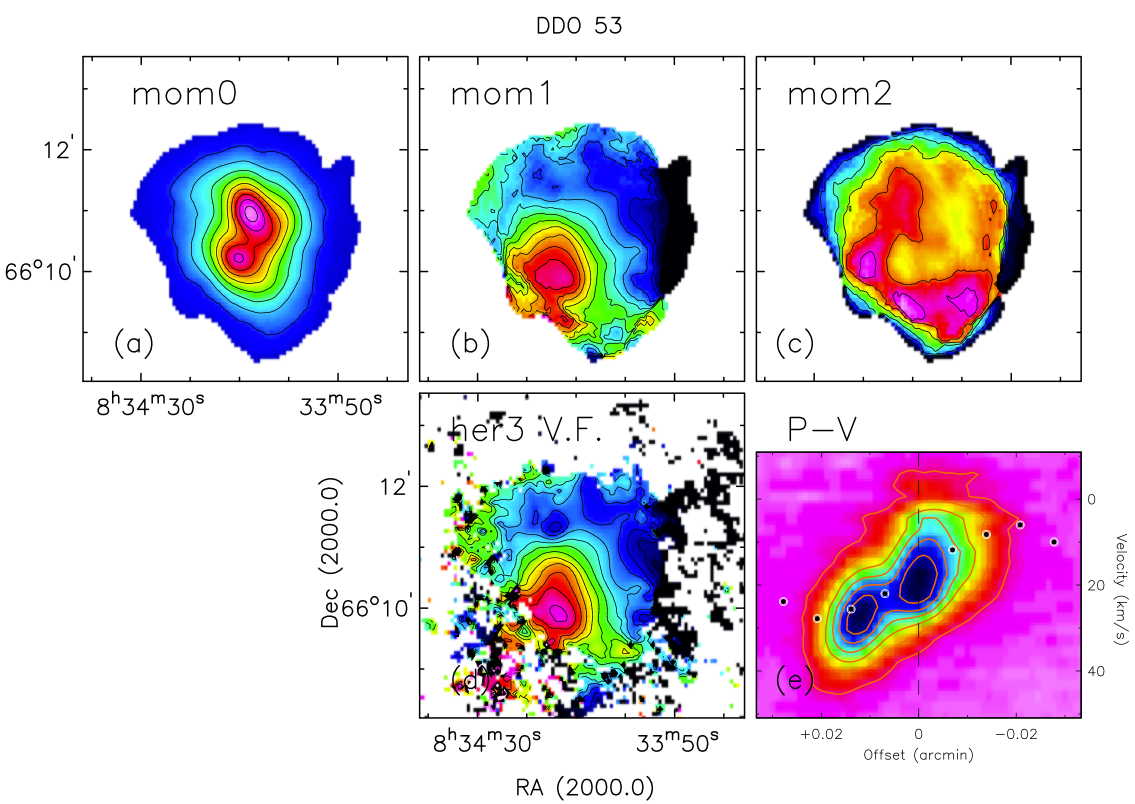}
\caption{Moment maps and a major-axis position-velocity diagram of DDO 53.
{\bf(a)} Integrated H{\sc i} map (moment 0). Contours run from 0 to 2 Jy beam$^{-1}$ \kms\ with a spacing of 0.2 Jy beam$^{-1}$ \kms.
The RA,DEC axes can be used to match the field of view with that in Figure \ref{fig-d53}
({\it middle right panel}).
{\bf(b)} Intensity-weighted mean velocity field (moment 1). Contours run from $-20$ \kms\ to $35$ \kms\, with a spacing of $2$\,\kms.
{\bf(c)} H{\sc i} velocity dispersion map (moment 2). Velocity contours run from $1$ to $17$\,\kms\,with a spacing of $2$\,\kms.
{\bf(d)} Hermite $h_{3}$ velocity field. Contours run from $-20$\, \kms\,to $35$\, \kms\,with a spacing of $2$\,\kms.
{\bf(e)} Position-velocity diagram taken along the average position angle ($155^{\circ}$) of the major axis.
Contours start at $+3\sigma$ in steps of $3\sigma$. The dashed lines indicate the systemic velocity and position
of the kinematic center derived from the tilted-ring analysis. Overplotted is the hermite velocity field rotation curve
corrected for the average inclination ($27^{\circ}$).
\label{fig-d53vf}}
\end{figure*}

\begin{figure*}
\includegraphics[angle=0,width=1.0\textwidth]{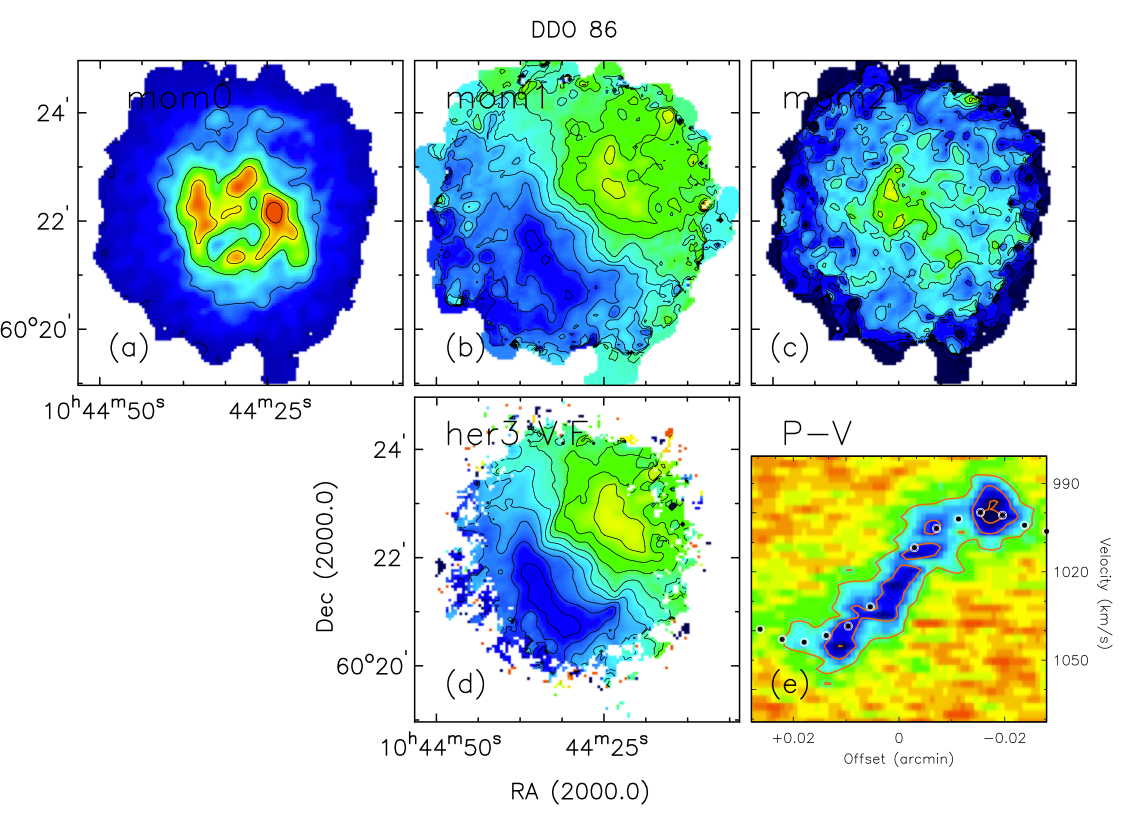}
\caption{Moment maps and a major-axis position-velocity diagram of DDO 86.
{\bf(a)} Integrated H{\sc i} map (moment 0). Contours run from 0 to 0.38 Jy\,beam$^{-1}$\,\kms\ with a spacing of 0.08 Jy\,beam$^{-1}$\,\kms.
The RA,DEC axes can be used to match the field of view with that in Figure \ref{fig-d86}
({\it middle right panel}).
{\bf(b)} Intensity-weighted mean velocity field (moment 1). Contours run from $1000$\,\kms\,to $1050$\,\kms\,with a spacing of $5$\,\kms.
{\bf(c)} H{\sc i} velocity dispersion map (moment 2). Velocity contours run from $1$ to $20$\,\kms\,with a spacing of $2$\,\kms.
{\bf(d)} Hermite $h_{3}$ velocity field. Contours run from $1000$\,\kms\,to $1050$\,\kms\,with a spacing of $5$\,\kms.
{\bf(e)} Position-velocity diagram taken along the average position angle ($317^{\circ}$) of the major axis.
Contours start at $+3\sigma$ in steps of $3\sigma$. The dashed lines indicate the systemic velocity and position
of the kinematic center derived from the tilted-ring analysis. Overplotted is the hermite velocity field rotation curve
corrected for the average inclination ($32^{\circ}$).
\label{fig-d86vf}}
\end{figure*}

\begin{figure*}
\includegraphics[angle=0,width=1.0\textwidth]{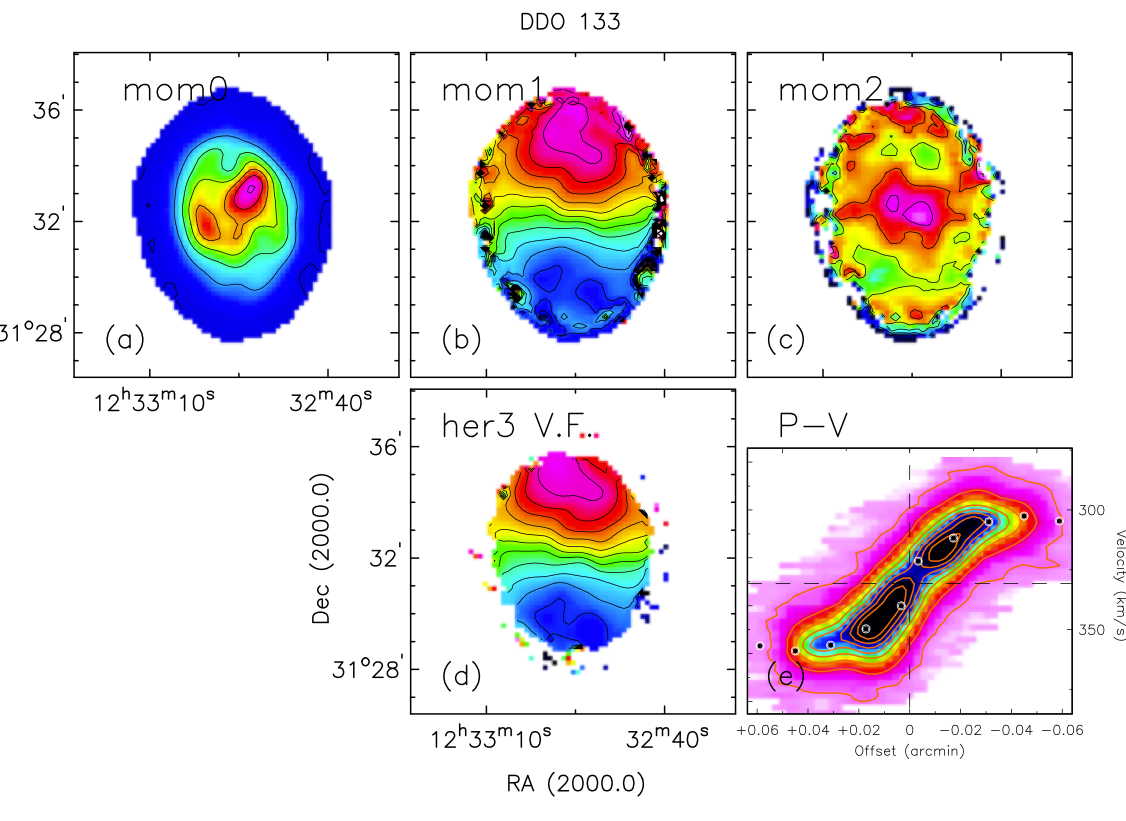}
\caption{Moment maps and a major-axis position-velocity diagram of DDO 133.
{\bf(a)} Integrated H{\sc i} map (moment 0). Contours run from 0 to 3 Jy\,beam$^{-1}$\,\kms\ with a spacing of 0.5 Jy\,beam$^{-1}$\,\kms.
The RA,DEC axes can be used to match the field of view with that in Figure \ref{fig-d133}
({\it middle right panel}).
{\bf(b)} Intensity-weighted mean velocity field (moment 1). Contours run from $280$\,\kms\,to $400$\,\kms\,with a spacing of $5$\,\kms.
{\bf(c)} H{\sc i} velocity dispersion map (moment 2). Velocity contours run from $1$ to $20$\,\kms\,with a spacing of $2$\,\kms.
{\bf(d)} Hermite $h_{3}$ velocity field. Contours run from $280$\,\kms\,to $400$\,\kms\,with a spacing of $5$\,\kms.
{\bf(e)} Position-velocity diagram taken along the average position angle ($358^{\circ}$) of the major axis.
Contours start at $+1\sigma$ in steps of $3\sigma$. The dashed lines indicate the systemic velocity and position
of the kinematic center derived from the tilted-ring analysis. Overplotted is the hermite velocity field rotation curve
corrected for the average inclination ($46^{\circ}$).
\label{fig-d133vf}}
\end{figure*}

\begin{figure*}
\includegraphics[angle=0,width=1.0\textwidth]{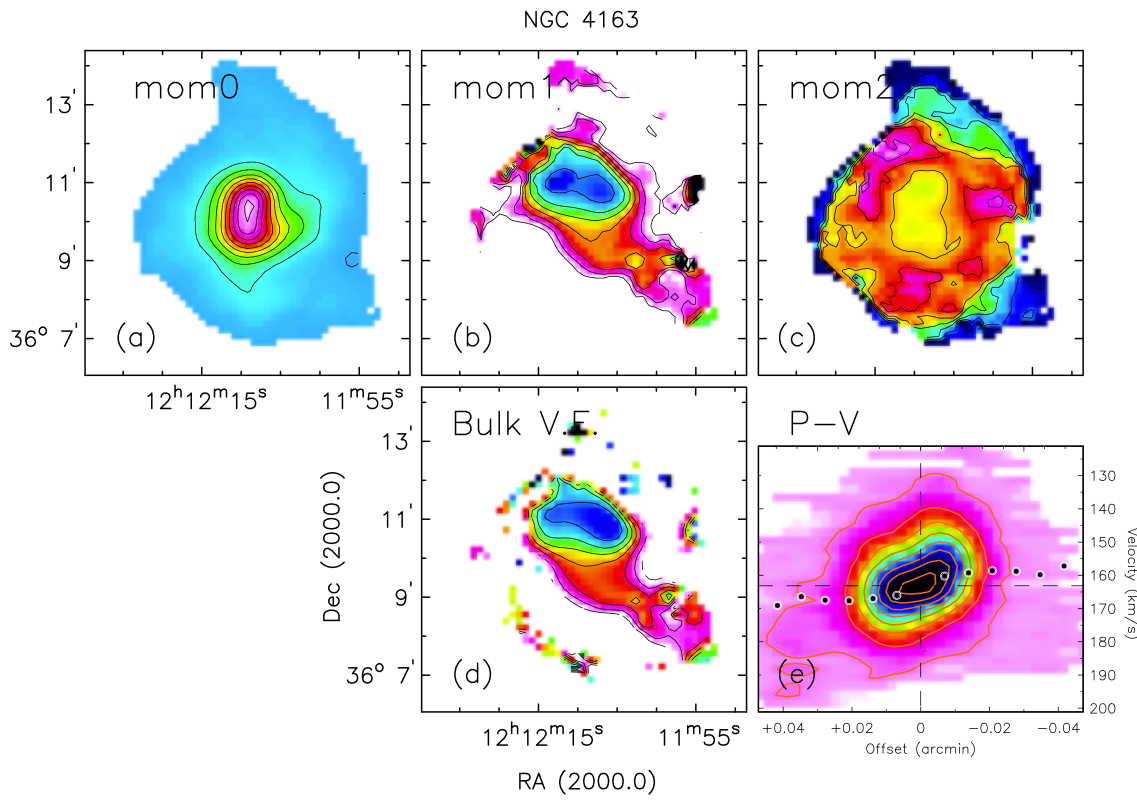}
\caption{Moment maps and a major-axis position-velocity diagram of NGC 4163.
{\bf(a)} Integrated H{\sc i} map (moment 0). Contours run from 0 to 1.9 Jy\,beam$^{-1}$\,\kms\ with a spacing of 0.2 Jy\,beam$^{-1}$\,\kms.
The RA,DEC axes can be used to match the field of view with that in Figure \ref{fig-n4163}
({\it middle right panel}).
{\bf(b)} Intensity-weighted mean velocity field (moment 1). Contours run from $150$\,\kms\,to $170$\,\kms\,with a spacing of $2$\,\kms.
{\bf(c)} H{\sc i} velocity dispersion map (moment 2). Velocity contours run from $1$ to $17$\,\kms\,with a spacing of $2$\,\kms.
{\bf(d)} Bulk velocity field. Contours run from $150$\,\kms\,to $170$\,\kms\,with a spacing of $2$\,\kms.
{\bf(e)} Position-velocity diagram taken along the average position angle ($15^{\circ}$) of the major axis.
Contours start at $+1\sigma$ in steps of $3\sigma$. The dashed lines indicate the systemic velocity and position
of the kinematic center derived from the tilted-ring analysis. Overplotted is the hermite velocity field rotation curve
corrected for the average inclination ($48^{\circ}$).
\label{fig-n4163vf}}
\end{figure*}

\begin{figure*}
\includegraphics[angle=0,width=1.0\textwidth]{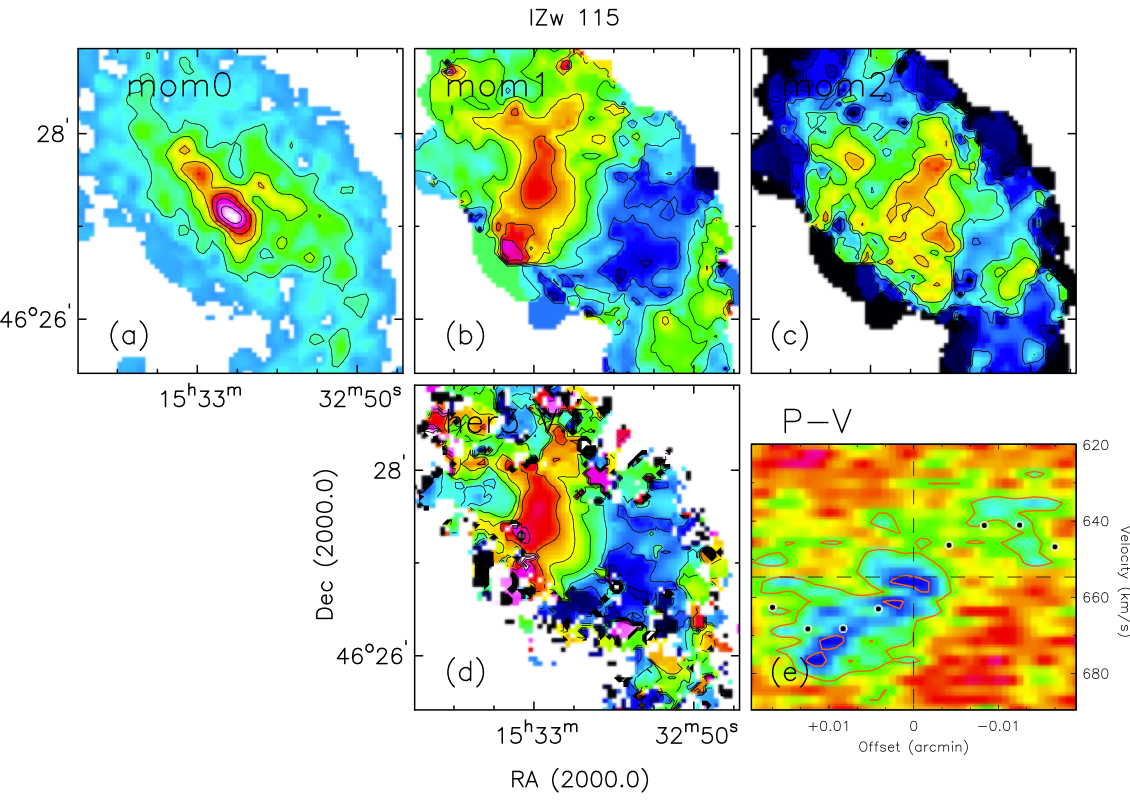}
\caption{Moment maps and a major-axis position-velocity diagram of IZw 115.
{\bf(a)} Integrated H{\sc i} map (moment 0). Contours run from 0 to 0.2 Jy\,beam$^{-1}$\,\kms\ with a spacing of 0.03 Jy\,beam$^{-1}$\,\kms.
The RA,DEC axes can be used to match the field of view with that in Figure \ref{fig-izw115}
({\it middle right panel}).
{\bf(b)} Intensity-weighted mean velocity field (moment 1). Contours run from $600$\,\kms\,to $700$\,\kms\,with a spacing of $5$\,\kms.
{\bf(c)} H{\sc i} velocity dispersion map (moment 2). Velocity contours run from $1$ to $15$\,\kms\,with a spacing of $2$\,\kms.
{\bf(d)} Hermite $h_{3}$ velocity field. Contours run from $600$\,\kms\,to $700$\,\kms\,with a spacing of $5$\,\kms.
{\bf(e)} Position-velocity diagram taken along the average position angle ($51^{\circ}$) of the major axis.
Contours start at $+3\sigma$ in steps of $3\sigma$. The dashed lines indicate the systemic velocity and position
of the kinematic center derived from the tilted-ring analysis. Overplotted is the hermite velocity field rotation curve
corrected for the average inclination ($61^{\circ}$).
\label{fig-izw115vf}}
\end{figure*}

\begin{figure}
\epsscale{1.0}
\includegraphics[angle=0,width=1.0\textwidth]{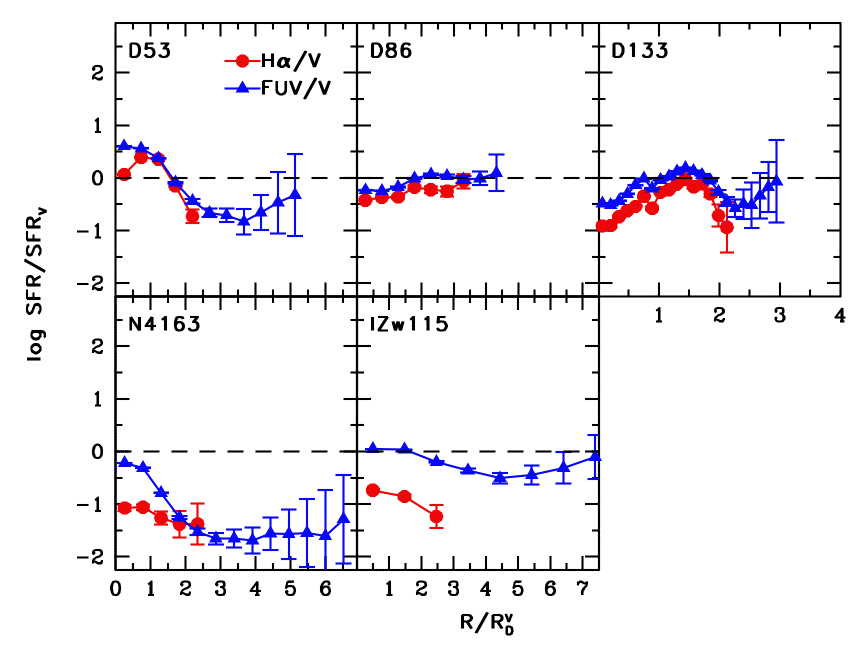}
\caption{Azimuthally-averaged star formation rates (SFRs) determined from the FUV
luminosity $SFR_{FUV}$ ({\it blue}) and from the H$\alpha$ $SFR_{H\alpha}$ ({\it red})
divided by $SFR_{V}$ for the 5 galaxies in our sample.
$SFR_{V}$ is determined from the mass
in starsÑcalculated from $M_V$ and a stellar $M/L_V$ ratio that depends on ($B-V$)$_0$, and
assumes a constant SFR over 12 Gyr.
$SFR_{H\alpha}$ is determined from the H$\alpha$ luminosity.
$SFR_{H\alpha}$ tells us of the past 10 Myrs; $SFR_{FUV}$ integrates over
a longer time scale; and $SFR_{V}$ is most sensitive to the past 1 Gyr for constant SFRs.
The horizontal dashed lines mark equal SFRs, or
ratios of 1. The ratios are plotted as a function of the radius normalized to the disk scale
length measured from $V$-band surface brightness profiles $R^V_D$.
\label{fig-sfrrat}}
\end{figure}

To compare the stars and gas, in Figures \ref{fig-d53sb} through \ref{fig-izw115sb} we show the
azimuthally-averaged surface photometry plotted against radius
from the galaxy center. The top panel in the graph includes properties of the gas:
the gas surface density $\Sigma_{gas}$ and velocity dispersion of the rotational component $\sigma_{vel}$.
The second panel shows the optical and ultraviolet properties:
$V$, \ha, and NUV surface photometry.
The third and fourth panels show optical and UV colors: $(B-V)_0$ and $(FUV-NUV)_0$.
From comparison of \HI\ and optical images (Figures \ref{fig-d53} through Figure \ref{fig-izw115}),
we see that the \HI\ does not extend much further than the stars in DDO 53 and DDO 133.
In DDO 133, there are bright UV regions that extend almost to the edge of the observed \HI, and the
big UV-bright complexes are found roughly near the ends of the stellar bar.
In DDO 53, NGC 4163, and IZw 115 the recent star formation is concentrated to the center, and we do not find any UV-bright knots
in the outer disk.
In DDO 86  there is a central hole in the gas. Many of the UV knots are associated with clouds around this hole.

In Figure \ref{fig-nhisfr} we plot the azimuthally-averaged $\log \Sigma_{gas}$ against
$\log {\rm SFR}$ for the
three star formation measures: SFR$_{FUV}$, SFR$_{H\alpha}$, and SFR$_V$.
This is similar to plots by Bigiel \et\ (2010) for pixel-by-pixel distributions.
We see a general trend of higher SFR for higher gas density except for some radii in DDO 53 and NGC 4163.
Those two galaxies are also are most peculiar kinematically.

The Toomre (1964) model of a differentially rotating thin disk
appears to explain star formation in the inner disks of spirals (Kennicutt 1989).
In this model there is a critical gas density \sigcrit\ above which the disk is
unstable to ring-like perturbations in the radial direction.
Gas that is above \sigcrit\ will form star-forming clouds; gas below \sigcrit\
is stable to cloud formation.
The ratio of \siggas\  to \sigcrit\  is shown as a function of radius in
the five galaxies in the top panels of the graphs in Figures \ref{fig-d53sb} through \ref{fig-izw115sb}.
\siggas/\sigcrit\ is everywhere less than one.
The ratio is extremely low throughout DDO 53 ($<$0.03),
but that may be due to the extremely low rotation speed in that galaxy.
The ratio of \siggas/\sigcrit\  is low ($<$0.3), but roughly constant where it was measured
over the disk of IZw 115.
For the other three galaxies---DDO 86, DDO 133, and NGC 4163---\siggas/\sigcrit\
rises to a peak outside the center and then declines to the outer disk.
In DDO 133, for example, the NUV surface photometry extends to the
regime where \siggas/\sigcrit\ is 0.05, well below the value to permit star formation.
So, how is star formation proceeding at very low gas column densities?

One possibility is that clouds of high enough density for star formation form in outer gas disks through
turbulence compression (Elmegreen \& Hunter 2006).
Our \HI\ maps are of low spatial resolution, but
along the northern edge of DDO 86, for example, there appear to be two \HI\ clouds that stand out.
Both have peaks that are 6$\sigma$ above the surrounding gas.
At 10$^h$ 44$^m$ 32.24$^s$, 60\arcdeg\ 23\arcmin\ 55\arcsec\ and at
10$^h$ 44$^m$ 26.17$^s$, 60\arcdeg\ 24\arcmin\ 01\arcsec\  (J2000),
the eastern cloud has a mass of $7\pm3\times10^6$ M\solar, a diameter of order 2 kpc,
and a surface density of order 2 M\solar\ pc$^{-2}$.
The western cloud has a mass of $4\pm2\times10^6$ M\solar, a diameter $\leq$1.4 kpc,
and a surface density of order 1 M\solar\ pc$^{-2}$.
These surface densities are right at the limits where star formation is found
to occur in other galaxies: 2-5 M\solar\ pc$^{-2}$
(Gallagher \& Hunter 1984; Skillman 1987; Taylor \et\ 1994;
Meurer \et\ 1996; Hunter \et\ 1998; van Zee \et\ 1998; Hunter \et\ 2001),
but much lower than would be expected for significant quantities of star formation
(Bigiel \et\ 2010).
Thus, clouds like these could potentially form stars.
Yet these clouds, at a distance of 5 disk scale lengths (9.2 kpc) from the center of the galaxy,
are surrounded by low column density gas ($1\times10^{20}$ cm$^{-2}$).

\section{Discussion}

\subsection{Radial Profiles}

Deep $V$-band images of 4 dwarf irregular galaxies and one BCD galaxy
have been used to measure surface photometry down to $\sim29.5$ mag
arcsec$^{-2}$.  Deep $B$-band images were also used for 3 of these
galaxies. The surface brightness distributions in the $V$-band were found to
be exponential in radius out to 5 inner disk scale lengths ($R_{\rm
D}$) in two cases (DDO 53, DDO 86), $8R_{\rm D}$ in two other cases
(NGC 4163, IZw 115), and $3R_{\rm D}$ in one case (DDO 133). A clear
break marks the transition from an inner exponential profile to an
outer, steeper exponential profile in three cases. In two of these (DDO
53, DDO 86), the break occurs at $\sim3R_{\rm D}$, and in the third
(DDO 133) it occurs at $\sim1.8R_{\rm D}$. In two cases where the $B-V$
color information goes beyond the break, one (DDO 53) has a transition
in the color profile from a slight reddening in the inner disk (from $0.12\pm0.005$ to $0.36\pm0.03$, a difference of 0.24 mag)
to an increasing blueness (down to $-0.13\pm0.3$ at the last point shown, a difference of 0.49 mag) in the outer disk;
the other (DDO 133) has a relatively flat color profile over all (variations up, down, up, and down again
with radius covering an extreme of $0.47\pm0.005$ to $0.30\pm0.01$, a difference of 0.17 mag).

The remarkable regularity of these underlying exponential disks
contrasts with the irregularity observed in this Hubble morphology
class. Such regularity has been observed in the inner parts of dwarf
galaxies for a long time (Patterson \& Thuan 1996; Hunter \et\ 1998;
Bremnes \et\ 1998, 1999, 2000;  Hopp 1999; van Zee 2000; Taylor \et\
2005; Hunter \& Elmegreen 2006; Amor\'in \et\ 2007; but for exceptions
see, for example, Hunter \& Gallagher 1985; Hunter \& Plummer 1996;
Herrmann \et, in preparation), but now we see that it extends into the
extreme outer disk as well. Somehow the accumulated star formation and
radial mass migration over a Hubble time converts the irregular
activity at any one time into a standardized disk profile. This
conversion seems to be done without grand design spiral or bar torques
(only one case, DDO 133, is clearly barred and none have obvious
spirals, but see below), and without significant shear in \HI\ velocity
maps for at least the inner exponential parts. A lack of bars would
mean that the Hohl (1971) mechanism for the formation of exponential
profiles by bar and spiral torques is not operating. A lack of shear
would suggest that the Lin \& Pringle (1987) mechanism in which the
star formation rate is proportional to the shear viscosity is also not
operating.

{\it GALEX} ultraviolet observations were also examined for these galaxies.
In four cases, the NUV observations go out as far in the disk as the
deep $V$-band images. The NUV profiles are also exponential and have
about the same scale lengths as the $V$-band profiles. In two of the
three cases with exponential breaks in $V$-band, there is also a break in
the NUV profile at the same radius. In the one case with a $B-V$ color
change at the break radius (DDO 53), there is no evident break in the
NUV profile at all, even though it extends for one inner disk scale
length beyond the $V$ break. This lack of a NUV break in DDO 53 could be
related to the increasing blueness with radius seen in $B-V$: the
blueness could compensate for the steeper mass profile seen in the $V$-band
and make the NUV-band appear continuous.

The similarity between the NUV and $V$-band profiles introduces another
puzzle about the formation and evolution of these galaxies. The average
radial profile of the star formation rate over the last several tenths
of a Gyr, as observed in the NUV, follows the underlying mass
distribution established by star formation and radial migration over
the last Gyr or more, as observed in $V$. There was no radial mixing of
these disks on timescales longer than $\sim1$ Gyr because that would
significantly affect the color profiles. For example, the outer
exponential disk should be redder than the inner disk if it was formed
or augmented by spiral arm scattering of inner disk stars $\sim1$ Gyr
old or more (e.g., Ro\v skar et al. 2008). Such a color change has been
observed for spiral galaxies (Bakos, Trujillo \& Pohlen 2008), but in
DDO 133 there is no color change and in DDO 53 the outer disk is bluer.
Neither galaxy has spiral arms anyway.

Consider an example using the population synthesis models of Bruzual \&
Charlot (2003). For a Salpeter IMF (the results are essentially the
same for a Chabrier IMF), a burst of star formation older than
$10^{8.5}$ yrs has $NUV-V>2$ mag, and one older than 1 Gyr has
$NUV-V>3.5$ mag. Young stars have $NUV-V<-1$.  These values were
determined by integrating the model stellar population spectra over the
filter functions used in the two passbands.  A plot of them is shown in
Figure 21 of our previous paper (Hunter et al. 2010). If we took a
group of stars that formed $10^{9}$ years ago and moved it to a
different place in the disk, then that group would be 3.5 magnitudes
fainter in NUV than in V because of population fading.  This 3.5 mag
difference means that the displaced group may still stand out in V but
be nearly invisible in NUV. Then the two passbands will have different
intensity distributions. This difference of 3.5 mag is so large that
even with more continuous star formation than a burst, and with some
mixing of the group into other stars, there should still be a
significant difference between the NUV and V intensities of stars $>1$
Gyr old. Our observations do not show this difference, however. The
similarity between the NUV and V profiles in our galaxies implies that
there has not been much migration of stars from the inner disk to the
outer disk on timescales longer than $\sim1$ Gyr.

The observation of similar radial profiles in NUV and $V$-bands (and
also in $J$-band for at least the inner regions -- see Hunter \&
Elmegreen 2006) suggests that star formation occurred at a steady rate.
Leroy et al. (2008) also found that the star formation rate in dwarfs
was about constant over a Hubble time. In only one of our cases (NGC
4163) is the average star formation rate observed in the NUV different
from the average observed in the $V$-band by as much as a factor of 10
(the UV rate is less than the $V$ rate at mid-disk), but in this case,
the overall kinematics of the \HI\ is very peculiar, suggestive of a
major disruption in the recent past. Still, NGC 4163 has a near-perfect
exponential disk in the $V$-band out to 8 scale lengths with no break;
it is the NUV profile that deviates from exponential at mid-radius.

H$\alpha$ observations are available for all of these galaxies from our
previous surveys. The H$\alpha$ radial profiles extend out to only
$2R_{\rm D}$ or $3R_{\rm D}$. In the case with the shortest $V$-band
break radius (DDO 133), the H$\alpha$ extends beyond the breaks in $V$
and NUV and also has a break at the same radius. In all of the
galaxies, the H$\alpha$ profile is roughly similar to the NUV profile,
including local deviations from the exponential. This can be seen in
Figure \ref{fig-sfrratfuvha} where we plot the ratio
SFR$_{FUV}$/SFR$_{H\alpha}$. The implication is that current star
formation rates, seen in H$\alpha$, are about the same at each radius
as the average rate over the last few hundred Myr. The H$\alpha$ rates
today are also about the same, to within a factor of $\sim3$, as the
$V$-band long-term average rates in three of the galaxies (see Figure
\ref{fig-sfrrat}). The two galaxies where these rates differ the most
are NGC 4163, which has highly disturbed kinematics as mentioned above
(and a current rate in H$\alpha$ and the UV that is much lower than the
$V$-band average at mid-radius), and the BCD galaxy, IZw 115, which
also has a low $H\alpha$ compared to $V$ in the inner parts (although
the UV rate is the same as the $V$ rate there). IZw 115 is evidently
post-starburst.

\subsection{Hydrogen Gas}

\begin{figure}
\epsscale{1.0}
\includegraphics[angle=0,width=0.9\textwidth]{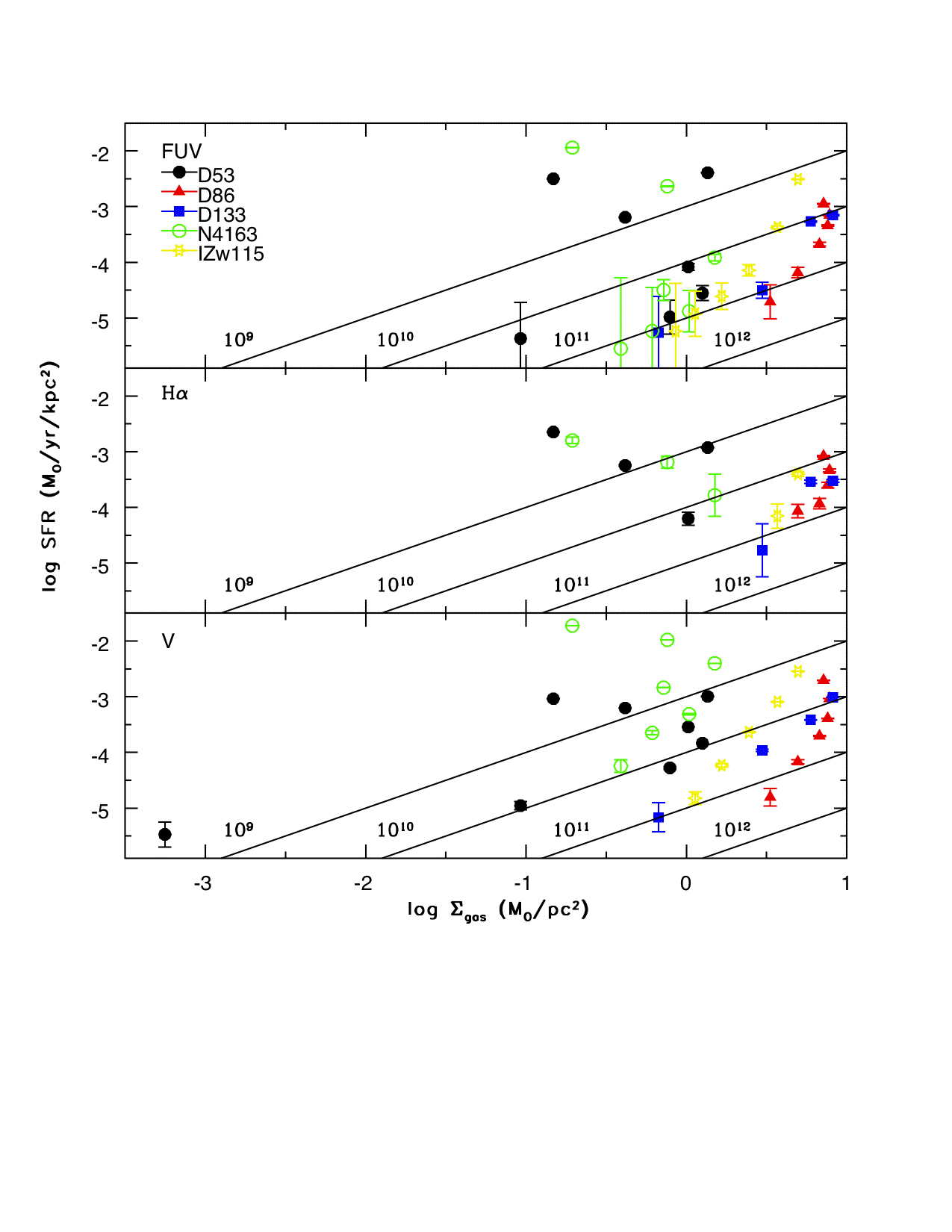}
\caption{Azimuthally-averaged $\log \Sigma_{gas}$ versus $\log SFR$ for
three star formation measures: SFR$_{FUV}$, SFR$_{H\alpha}$, and SFR$_V$.
The solid, slanted lines are lines of gas depletion times; the lines are labeled with the
gas depletion timescale in years.
\label{fig-nhisfr}}
\end{figure}

\begin{figure}
\epsscale{1.0} 
\includegraphics[angle=0,width=1.0\textwidth]{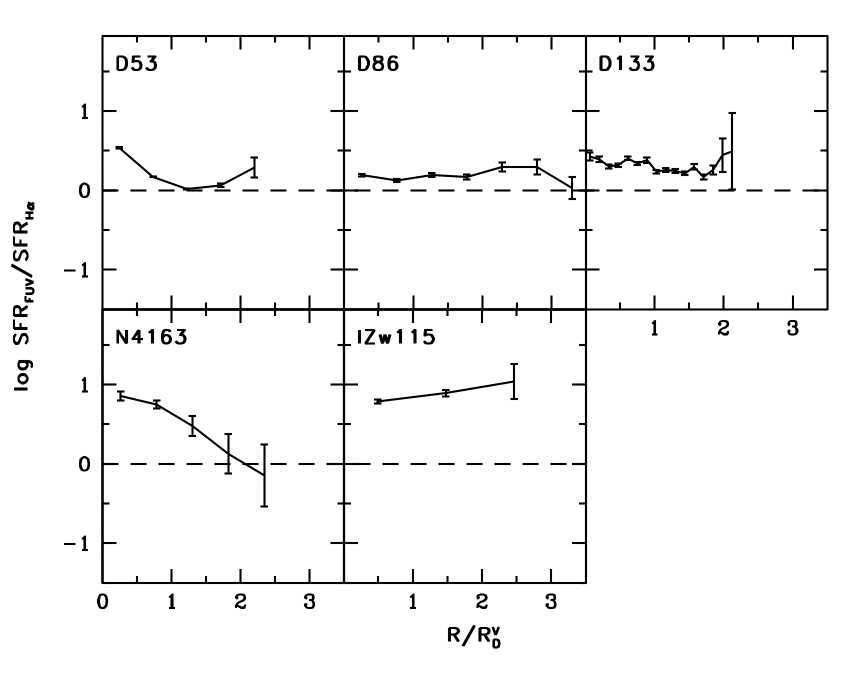}
\caption{Ratio of
azimuthally-averaged SFR determined from the FUV luminosity,
$SFR_{FUV}$, to that determined from the H$\alpha$, $SFR_{H\alpha}$.
$SFR_{H\alpha}$ tells us of the past 10 Myrs; $SFR_{FUV}$ integrates
over several hundred million years. The horizontal dashed lines mark
equal SFRs, or ratios of 1. The ratios are plotted as a function of the
radius normalized to the disk scale length measured from $V$-band
surface brightness profiles $R^V_D$. \label{fig-sfrratfuvha}}
\end{figure}

The final piece of evidence related to star formation and galactic
structure is the gas distribution. For this, we examined \HI\
observations taken at the VLA for this project. The \HI\ radial
profiles are smoothly declining but flatter than the stellar light
profiles in DDO 86, NGC 4163 and IZw 115, and about the same as the
stellar light in DDO 133. In DDO 86, there is an \HI\ hole in the
center and a ring around it that is not present in the optical. These
trends show up in the SFR-$\Sigma_{\rm gas}$ diagrams of Figure
\ref{fig-nhisfr} too.  DDO 86, DDO 133, and IZw 115 have steep
near-power-law relations between the $V$-band or FUV SFRs and the \HI\
column densities, with a slopes of $3\pm1$ to $4\pm1$. NGC 4163 and IZw
115 have a scattering of points with a somewhat more organized relation
for FUV than $V$-band SFRs. The near-power-laws are consistent with
other studies of star formation and \HI\ column density, which also
show steep slopes. Bigiel et al. (2010) got a slope of 1.7 for the
relation between SFR per unit area and \HI\ column density in faint
outer regions. Their plot of $\Sigma_{HI}$ versus $\Sigma_{SFR}$
for dwarfs in their Figure 8 is similar in shape to ours for the FUV
SFR in Figure \ref{fig-nhisfr}. Star formation follows unseen CO
(Bigiel et al. 2011), which is a small fraction of the \HI\ mass in
these faint regions (Schruba et al. 2011). The timescale for conversion
of \HI\ into stars is the ratio of the $x$ axis values to the $y$-axis
values in Figure \ref{fig-nhisfr}, or approximately $10^{11}$ years
(see also Bigiel et al. 2010).

The \HI\ distributions are clumpy with giant clouds that appear
associated with star formation. These clouds are most obvious in the
inner regions where the star formation rates are highest, although
faint \HI\ clouds can also be seen in the outer parts where past-average
star formation rates are very low.  Observable clouds
typically have masses of around $10^7\;M_\odot$ or larger. A few faint
clouds in the outer part of DDO 86 with masses of several
$\times10^6\;M_\odot$ were discussed as possible future sites of star
formation.  In general there are no spiral arms or systematic streaming
motions in \HI. The velocity fields reflect normal rotation curves
except for NGC 4163 and DDO 53.

The most extended \HI\ rotation curves are available for DDO 86, DDO 133,
and IZw 115. The first two have rising inner rotation curves that
flatten out at the break radius in the $V$-band profile. The third has a
rising rotation curve in the inner part and a falling rotation curve in
the outer part, with no break in the exponential profile for
comparison. There could be a small companion galaxy near IZw 115 that
affects the rotation curve in the outer part. The coincidence in two
cases between the turnover point in the \HI\ rotation curve and the break
point in the photometric radial profile is interesting but maybe not
significant. In our larger sample (Hunter \& Elmegreen 2006), there was
no correlation between the photometric break radius and the rotation
curve turnover radius: only one case had these features at the same
radius, 5 had the break at a larger radius than the turnover and 4 had
the break at a smaller radius than the turnover. In larger spiral
galaxies, the exponential break radii are typically at $\sim2-5R_{\rm
D}$ (Pohlen \& Trujillo 2006) and the turnover radii are at
$\sim1R_{\rm D}$, so there is no coincidence of radii for spirals
either.

The \HI\ rotation speeds in these dwarf galaxies are typically low: in
the three cases with extended rotation curves, they peak at 40 km
s$^{-1}$ for DDO 86 and DDO 133, and at 15 km s$^{-1}$ for IZw 115. The
\HI\ velocity dispersions are approximately constant with radius at
around 10 km s$^{-1}$.  For IZw 115, the average rotation speed is
comparable to the velocity dispersion inside $3R_{\rm D}$ and possibly
in the outer regions too where the rotation curve drops.  At the peak
of the rotation curve in IZw 115, the ratio of the rotation speed to
dispersion is remarkably low, $\sim1.4$.  Such slow rotation gives
IZw 115 very little dynamical integrity, yet it has a near-perfect
exponential disk for $8R_{\rm D}$.

Slowly rotating galaxies should be easily perturbed by interactions with
other galaxies and dark matter concentrations, by clumping
instabilities in their disks, and by gas blowout during star bursts.
This sensitivity to asymmetric forcing and rapidly-changing potentials
makes the long-range exponential structure of the disks difficult to
understand, unless it results from such forcings.  The uniform rotation
observed in 2 out of 5 cases suggests that their recent lives have been
quiescent, at least for the last several rotation periods in their
outer regions. The rotation period at the outer radius of detection for
the \HI\ rotation curve in DDO 133 is 0.9 Gyr. For DDO 86, it is 1.9 Gyr.
The SFR ratios shown in Figure \ref{fig-sfrrat} are roughly consistent with this long term
quiescence.

\begin{figure}
\epsscale{1.0}
\includegraphics[angle=0,width=0.8\textwidth]{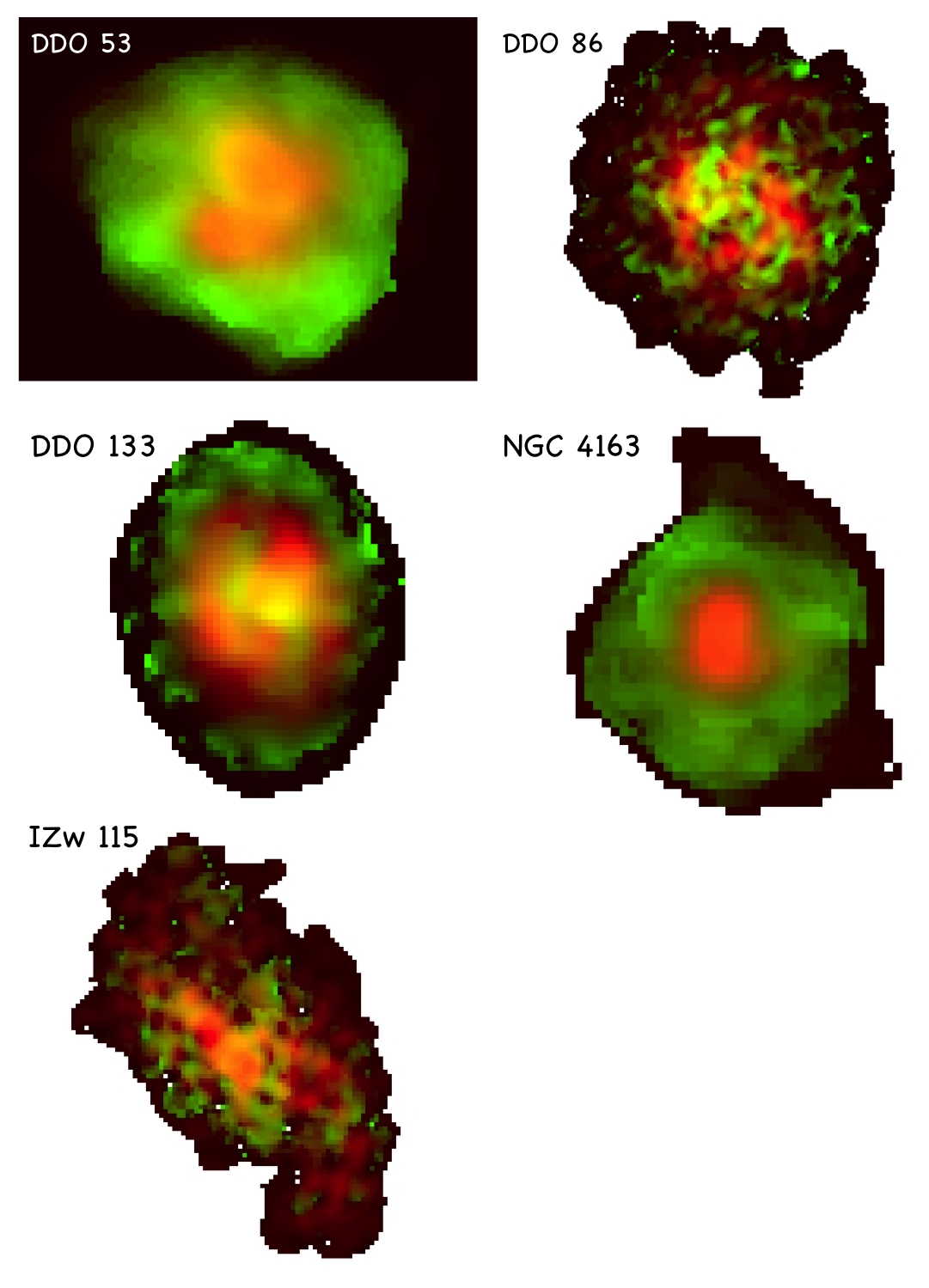}
\caption{Integrated \protect\HI\ maps (``mom0'') ({\it red}) superposed with velocity dispersion
maps (``mom2'') ({\it green}).
\label{fig-m0m2}}
\end{figure}

The maps of velocity dispersion, ``mom2'', in Figures \ref{fig-d53vf}
to \ref{fig-izw115vf} show an inverse correlation with the maps of
column density, ``mom0,'' in the sense that higher column densities
correspond to lower velocity dispersions.  We found this before for NGC
2366 (Hunter et al. 2001).  Figure \ref{fig-m0m2} shows this anti-correlation
better. Each panel shows a map of mom0 in red and mom2 in green.
Overlapping regions turn out yellow. The maps show mostly red and green
regions with only a little yellow. This means that the dispersion is
high where the column density is low, and vice versa. An
anticorrelation is roughly consistent with pressure uniformity in the
sense that low density regions have higher turbulent speeds. Piontek \&
Ostriker (2005) found this in simulations of the magneto-rotational
instability.

Trujillo \& Bakos (2010) looked for a correlation between possible
inflection points in the \HI\ velocity dispersion and exponential break
radii in the disks of local spiral galaxies. They were considering that
the breaks could result from gas disk flaring, but found nothing. We
see nothing like that here either.

\subsection{Gravitational Stability}

The \HI\ appears to be gravitationally stable by itself. The ratios of
the gas surface densities to the Toomre (1964) threshold densities are
0.6 or less in DDO 86 and DDO 133, 0.1 or less in NGC 4163, and 0.3 in
IZw 115. There are stars mixed with the gas, however, and the stellar
velocity dispersions are probably not much larger than the gaseous
dispersions, considering the low rotation speeds. This means that stars
contribute to the instability in approximate proportion to their mass
(Jog \& Solomon 1984).

We have estimated the mass surface densities of stars in our galaxies
using the $V$-band radial profiles and the $B-V$ colors, where available.
The calibration comes from $M/L$ tables in Bell \& de Jong (2000).  We
find that the stellar mass surface densities in the outer parts of our
galaxies are comparable to the gas surface densities, or slightly
smaller, to within a factor of a few. This can also be derived from the
star formation rates given in Table \ref{tab-sfrs}, which are
$\sim10^{-4.8}\;M_\odot$ yr$^{-1}$ kpc$^{-2}$ in the $V$-band at
$\mu_{\rm V}=29.5$ mag arcsec$^{-2}$. Multiplying this by $10^{10}$
years and converting to pc$^{-2}$, the result is a total stellar
surface density today of $\sim0.2\;M_\odot$ pc$^{-2}$ in the far outer
parts. Figures \ref{fig-d53sb} to \ref{fig-izw115sb} give $\Sigma_{\rm gas}$
there, and the average is about $1\;M_\odot$.  Thus gravitational
forcing from stars may be $\sim20$\% of that from gas during the growth
of an instability, leading to an upward correction to the surface
density ratios by $\sim20$\%.  The correction would be larger in the
inner regions, where the stellar contribution is larger. These
corrections make DDO 86 and DDO 133 only slightly stable, which is
typical for galaxies. For example,  Leroy et al. (2008) showed that the
gas+star stability parameter equals about 1.5 for a wide range of radii
inside dwarfs and spirals. The corresponding ratio of column density to
critical column density is the inverse of this, 0.6. This is not much
different than it is for DDO 86 and DDO 133.

NGC 4163 is not a good case for this stability analysis because the
rotation curve is peculiar (\S 4.3). IZw 115 could also have a peculiar
rotation curve because of the companion galaxy, but even in the inner
parts where the rotation curve is rising, the \HI\ column density is
less than the critical value by a factor of 5. These inner parts are
probably post-starburst, however, so they could well be very stable.

The situation with instabilities is more complicated than this because
all of these galaxies have relatively thick disks, as inferred from the
high ratios of HI velocity dispersion to rotation speed. Thickness $H$
decreases the in-plane accelerations from self-gravity at wavenumber
$k$ by a factor of $\sim(1+kH)^{-1}$ (Vandervoort 1970), which should
be $\sim0.5-0.7$ in our case.  This lowers the effective stability
ratio $\Sigma/\Sigma^\prime_{\rm crit}$ by the same factor, since
$\Sigma^\prime_{\rm crit}=\kappa\sigma(1+kH)/\pi G$ for $Q^\prime=1$
with thickness corrections.  Here, $\sigma$ is the velocity dispersion,
$\Sigma$ is the mass column density, and $Q^\prime=\kappa\sigma(1+kH) /
\pi G \Sigma$ is the thickness-corrected Toomre $Q$ parameter for
epicyclic frequency $\kappa$ (Elmegreen 2011). The ratios derived above
for gas+stars in DDO 86 and DDO 133 without thickness corrections are
comparable to those given by Leroy et al. (2008) for dwarf and spiral
galaxies without thickness corrections. Therefore both are likely to
have comparable stability ratios when thickness is included, which
means both are slightly more stable than without the thickness
correction. The ratios for the outer parts of dwarfs would still be
smaller than they are for spirals.

Gravitational instabilities in dwarf irregular galaxies should look
different than they do in large spiral galaxies. The characteristic
length for spirals or clumps that form in a gravitational instability
lies in the range $\sigma^2/\left(\pi G \Sigma^\prime\right)\left(1\pm
\left[1-Q^2\right] \right)^{-1}$. This is centered at
$\sigma^2/\left(\pi G \Sigma^\prime\right)$, which should be compared
with the disk scale length. In dwarfs with comparable velocity
dispersions for the gas and stars, we can use the total mass column
density for $\Sigma^\prime$ and include both components this way. In
the outer parts of dwarf galaxies, $\Sigma\sim10\;M_\odot$ pc$^{-2}$,
$\Sigma^\prime\sim0.5\Sigma$, and $\sigma\sim10$ km s$^{-1}$, so
$\sigma^2/\pi G \Sigma^\prime =1.4$ kpc. For spiral galaxies, the
length scale is about the same because the velocity dispersion for the
dominant stars is higher by a factor of $\sim3$, and the mass column
density is higher by a factor of $\sim10$. Dwarf galaxies have smaller
exponential scale lengths than spirals, however. Three of the dwarfs
studied here have $R_{\rm D}<0.7$ kpc and two others have $R_{\rm
D}\sim2$ kpc (Table \ref{tab-disk}). In spiral galaxies, $R_{\rm D}$
ranges between $\sim1$ and 5 kpc (van der Kruit 1987), which is larger
than in dwarfs by a factor of 2 or more.  This difference means that
the characteristic length of a significantly gravitating structure in a
large spiral galaxy readily fits into the disk, allowing coherent
structures like spiral waves to form. Most dwarfs are too small for
this.  In addition, the bright parts of dwarf galaxy disks occur in the
rising parts of their rotation curves, but the main parts of spiral
galaxy disks are in the flat parts of their rotation curves.
Gravitational perturbations therefore tend to grow as unsheared clumps
in the dwarfs, but they produce spiral wakes and long sheared arms in
spirals.

Some dwarf galaxies have spirals, but this situation is rare (Schombert
et al. 1995). A dwarf elliptical in the Virgo cluster, IC 3328, has a
weak spiral, but it may be confined to a thin sub-component of the main
stellar disk where the gravitational length is short (Jerjen et al.
2000; De Rijcke et al. 2003; Lisker \& Fuchs 2009). Other dwarf spirals
in Virgo were observed by Barazza et al. (2002).

The same considerations apply to the far-outer parts of spiral
galaxies, where $\Sigma$ is comparable to that in dwarfs. Then the
characteristic length for instabilities may be 10 times higher than it
is in the inner spiral disk, and locally much larger than the scale
length. Thus the outer parts of spiral disks should have difficulty
generating self-gravitating waves on their own, like dwarf galaxies.
However, the outer disks of spiral galaxies can have spiral waves that
propagate there from the inner disk (Bertin \& Amorisco 2010) or were
resonantly excited by perturbations from the inner disk (Debattista et
al. 2006). Dwarf galaxies have neither source of waves and should
produce only clumps of self-gravitating gas.

Disk stability is also relevant to the origin of double exponential
profiles, according to the model by Debattista et al. (2006). This
model involves angular momentum transfer around bars and spirals.
Debattista et al. (2006) showed that the inner parts of barred galaxies
with single exponential profiles get flatter with time, out to the
outer Lindblad resonance of the spiral that is driven by the bar. The
original exponential disk then shows up as the outer steep part of the
double exponential, and the bar-modified part is the inner flat region.
The degree of flattening depended on $Q$ in these models. Disks that
were more gravitational stable, with higher $Q$, did not evolve as much
as less stable disks. This result suggests that dwarf galaxies that are
weakly self-gravitating may not be responsive to bar torques that
modify a single exponential into a double exponential.

Our galaxies with the highest ratios of $\Sigma_{\rm gas}/\Sigma_{\rm
crit}$, namely DDO 86 and DDO 133, also have the most obvious breaks in
their exponential profiles. DDO 133 is a barred galaxy too. However,
the break radius in DDO 133 is at the end of the bar and not the end of
a spiral as in the Debattista et al. model. Also in DDO 86, the star
formation region and inner bright core in Figure \ref{fig-d86} is
elongated in the east-west direction, reminiscent of a bar.  The end of
this elongation corresponds to the break radius. Similarly in DDO 53,
which is the only other case here with a double exponential, there is
an elongation of the main disk star-forming region along a position
angle of $\sim150^\circ$, which is transverse to the major axis. The
deprojected length of this elongation is comparable to the break
radius. There is not enough \HI\ data to determine $\Sigma_{\rm
gas}/\Sigma_{\rm crit}$ for this galaxy.  These observations suggest
that the inner exponentials of our galaxies were made flatter than the
outer exponentials by mass transfer connected with a bar. There are no
spirals here, so the bar may be doing here what the spirals did in
Debattista et al. (2006).

We looked for a correlation between photometric breaks and the presence
of bars in our larger sample (Hunter \& Elmegreen 2006).  Out of 22
dwarf galaxies with double exponentials of the type discussed here, 8
have some evidence for a bar (even if they are not classified as Hubble
bar types), and the rest do not. DDO 133, studied here, was one of the
barred cases with a double exponential recognized in that previous
study. Among the barred dwarfs, the ratio of the break radius to the
end of bar radius is $1.8\pm0.6$ on average. Since there are breaks
without bars and bars without breaks, we do not see a clear correlation
between the two at this time.

\subsection{Two Peculiarities}

We summarize this discussion by highlighting two peculiar things about
the present observations: the extraordinarily regular exponential disks
in dwarf irregular galaxies, and the continuity of star formation
throughout these disks and over time. Exponential disks are made in
part by cosmological infall (Freeman 1970; Fall \& Efstathiou 1980; van
der Kruit 1987) and subsequent dynamical torques from self-gravitating
disk structures. There is no regularity to these structures in dwarfs
and no systematic presence of bars or spirals. There are only clumps of
star formation and maybe some old disk stars mixed in with those
clumps. It is conceivable that the clumps of star formation make or
maintain the exponential disk, as in numerical simulations of equally
clumpy galaxies in the early universe (Bournaud et al. 2007). Or, there
could be bar-like dark matter halos (Bekki \& Freeman 2002; Tutukov \&
Fedorova 2006) that redistribute angular momentum, while also driving
in gas to fuel occasional low-level bursts (Hunter \& Elmegreen 2004).

As for star formation itself, this complex process involving mostly gas
seems to follow the underlying exponential disk of old stars better
than one would expect from such a stochastic process. Leroy et al.
(2008) found the same thing for dwarfs and the outer parts of spiral
galaxies. Molecular cloud distributions follow the stellar
exponential even when the molecules are an extremely low fraction of
the gas mass (Schruba et al.\ 2011). How do the molecules know to do
this? In the current paradigm, star formation follows only the
molecules, and molecular abundances depend mostly on the pressure
(e.g., Bigiel et al. 2008), but why does this end up reinforcing the
underlying exponential disk? There is a purely radial dependence to the
cloud and star formation rates, in addition to a gas dependence, as in
the dynamical law of star formation, where the rate is determined by
the orbit time (Kennicutt 1998). Bigiel et al. (2010, Figure 7) also
found that for a given \HI\ column density, the star formation rate
decreases with increasing radius in spiral and dwarf galaxies.  The
problem with this is that orbital timescales usually vary with radius
as a power law, not an exponential.

Ostriker et al. (2010) proposed that the star formation rate scales
with the total gas surface density multiplied by the square root of the
midplane density from stars and dark matter. They get good agreement
between this prediction and the radial profiles of star formation in
several spiral galaxies.  The present results suggest that if this
quantity applies to dwarf irregulars, then it should also be
proportional to the column density of existing stars so that the color
profile does not change much over at least a Gyr.  In fact they find
the star formation rate per unit gas mass proportional to the surface
density of existing stars and dark matter in the case where the
vertical velocity dispersion is constant (because then the midplane
density scales with the square of the stellar and dark matter column
density).  This matches our result fairly well because the gas column
density varies weakly with radius.

Another model is that in highly stable regions, molecular cloud
formation and star formation is somehow triggered by field stars.
Perhaps Type Ia supernovae are important, or the turbulence that is
driven by these supernovae and other field-star pressures. Star
formation would then follow the underlying stellar density as long as
there is enough gas to compress. Since the rate is approximately
constant over time, sequential triggering by young star pressures could
also be important.  Then star formation would move around as
illustrated by Gerola et al. (1980).  A way to check these models would
be to find triggering sources near young stars.  Also, the star
formation rate should be more tightly correlated with the stellar
column density than with the gas column density, as long as there is
enough gas to be triggered.

The break radii in exponential disks could be a clue to these processes
if the properties of the gas and cosmic rays near regions of star
formation were found to differ inside and outside of this radius. In
the present sample of galaxies, we found no clear correlation between
the existence of a break and other properties, such as bars, spirals,
color jumps, or the turnover in the rotation curve. We found no such
correlations in our bigger sample either (Hunter \& Elmegreen 2006).

\section{Summary}

We present ultra-deep $V$, $B$, and {\it GALEX} UV images obtained in order to detect and measure
the extreme outer stellar disks of a sample of 4 dIm galaxies and one BCD galaxy.
Our $V$-band surface photometry extends to 29.5 magnitudes arcsec$^{-2}$.
We convert our new FUV and $V$-band photometry, along with \ha\ photometry obtained in a larger survey,
into radial star formation rate profiles
that are sensitive to different timescales---from 10 Myrs to the lifetime of the galaxy.
We also present \HI-line emission data and compare
the stellar distributions, surface brightness profiles, and star formation rate profiles to \HI-line emission maps,
gas surface density profiles, and gas kinematics. Our data lead us to two remarkable general observations.

First, the exponential disk in these dwarf irregular galaxies are extraordinarily regular.
We observe that the stellar disks continue to decline exponentially as far as our measurements extend.
In addition, as is common in outer spiral and dwarf disks,
the slope of the $V$-band surface brightness profiles change abruptly in the outer disks
in most of our systems. In three of our galaxies the outer profile becomes steeper, but
the break in the $V$-band surface brightness profile is mimicked in the NUV profile for two of these systems.
The galaxies are lumpy with star formation and the
\HI\ distributions and kinematics have significant non-ordered motions.
Yet, the regularity of the exponential profiles in spite of these peculiarities
indicates that the sporadic processes that have built the
disks---star formation, radial movement of stars, and perhaps even perturbations from the outside---have,
nevertheless, conspired to produce standard disk profiles.

Second, there is a remarkable continuity of star formation throughout
these disks over time. In four out of five of our galaxies the star
formation rate in the outer disk measured from the FUV tracks that
determined from the $V$-band, to within factors of a few ($<$5),
although the FUV-derived rate reflects a shorter timescale than the
$V$-band derived rate. This requires star formation at a fairly steady
rate over the galaxy's lifetime. But how the galaxy achieves this is
not clear. The \HI\ surface density profiles generally decline with
radius more shallowly than the stellar light and, hence, star formation
rate profiles. The star formation instead appears to follow the older
stars. There is a general correlation between azimuthally-averaged
$\log \Sigma_{gas}$ and $\log {\rm SFR}$, similar to what is seen by
others, except for some radii in the two galaxies with the disturbed
kinematics. The gas is marginally gravitationally stable against
collapse into clouds in these galaxies.

Outer stellar disks are being traced to ever fainter levels in dwarf galaxies, and
these observations are challenging our concepts of star formation and disk growth over cosmic times.
Clearly, understanding the processes that mold outer disks,
as well as understanding how stars form in the extreme environment of outer disks,
are important to our understanding of galaxy formation and evolution.

\acknowledgments
Support for this work was provided to DAH by the Lowell Observatory
Research Fund, by grant AST-0204922 from
the National Science Foundation, and by grant NNX08AL66G from NASA/{\it GALEX}.
Funding to BGE was provided by NSF grant AST-0707426.
SHOH acknowledges financial support from the South African Square Kilometre Array Project.
MR and NW are grateful for participation in the Northern Arizona University Research
Experiences for Undergraduates program in the summers of 2008 and 2009, respectively.
This program is run by Kathy Eastwood and funded by the National Science Foundation
through grant AST-0453611.
We are also grateful to the VLA, KPNO, and {\it GALEX} personnel
for access to the superb instrumentation that made this project possible and all of the effort that
goes into making such facilities work.

Facilities: \facility{VLA}, \facility{KPNO},  \facility{GALEX}  

\clearpage

\clearpage

\clearpage

\clearpage

\clearpage

\clearpage

\clearpage

\clearpage

\clearpage

\clearpage

\clearpage

\clearpage

\clearpage

\clearpage

\clearpage

\clearpage

\clearpage

\clearpage

\clearpage

\clearpage

\clearpage

\clearpage

\begin{deluxetable}{lllrrc}
\tabletypesize{\scriptsize}
\rotate
\tablecaption{Galaxy Sample \label{tab-sample}}
\tablehead{
\colhead{} & \colhead{} & \colhead{} & \colhead{D} & \colhead{}
& \colhead{log SFR$_D$\tablenotemark{e}} \\
\colhead{Galaxy} & \colhead{Other Names\tablenotemark{a}}
& \colhead{Type\tablenotemark{b}}
& \colhead{(Mpc)\tablenotemark{c}}
& \colhead{E(B$-$V)$_f$\tablenotemark{d}}
& \colhead{(M\protect\solar/yr/kpc$^2$)}
}
\startdata
DDO 53   & PGC 24050, UGC 4459, A0829$+$66, VIIZw238 & Im & 3.6\x & 0.03\xx & $-2.50$ \\
DDO 86   & PGC 32048, UGC 5846, A1041$+$60, Mailyan 057& Im & 16.9\x & 0.00\xx & $-2.74$ \\
DDO 133  & PGC 41636, UGC 7698, A1230$+$31            & Im  &  6.1\x & 0.00\xx & $-2.93$ \\
NGC 4163 & PGC 38881, UGC 7199, KUG 1209$+$364 & Im & 2.8\x & 0.00\xx & $-2.43$ \\
IZw 115  & PGC 55381, UGC 9893, VV 720, A1531$+$46, KIG 686& BCD & 12.3\x & 0.01\xx & $-2.87$ \\
\enddata
\tablenotetext{a}{Selected alternate identifications
obtained from NED.}
\tablenotetext{b}{Morphological Hubble types are from de Vaucouleurs et al.\ (1991). If no type is
given there, we have used types given by NED.}
\tablenotetext{c}{References for distances can be found in Hunter \& Elmegreen (2006).}
\tablenotetext{d}{Foreground Milky Way reddening from Burstein \& Heiles (1984).}
\tablenotetext{e}{Integrated star formation rate derived from the H$\alpha$ luminosity, divided by $\pi R_D^2$, where $R_D$ is the
disk scale length determined from the inner $V$-band surface brightness profile  from Hunter \& Elmegreen (2004).}
\end{deluxetable}

\clearpage

\begin{deluxetable}{lccccccccccrr}
\tabletypesize{\scriptsize}
\setlength{\tabcolsep}{0.02in}
\rotate
\tablecaption{Optical Observations and Photometry Parameters \label{tab-obs}}
\tablehead{
\colhead{} & \colhead{} & \colhead{} & \colhead{} & \colhead{} & \colhead{}
& \colhead{}
& \colhead{}
& \multicolumn{5}{c}{Ellipse photometry parameters\tablenotemark{d}} \\
\cline{9-13}
\colhead{} & \colhead{Date} & \colhead{} &\colhead{}
& \colhead{Exposures}
& \colhead{Seeing\tablenotemark{b}}
& \colhead{Scale}
& \colhead{Calib rms\tablenotemark{c}}
& \colhead{P.A.}
& \colhead{} & \colhead{Step} & \multicolumn{2}{c}{Center (J2000)} \\
\colhead{Galaxy}
& \colhead{(month/year)}
& \colhead{Tel.\tablenotemark{a}}
& \colhead{Filters}
& \colhead{(s)}
& \colhead{(arcsec)}
& \colhead{(arcsec)}
& \colhead{(mag)}
& \colhead{(deg)}
& \colhead{{\it b/a}} & \colhead{(arcsec)}
& \colhead{R.A.} & \colhead{Decl.}
}
\startdata
DDO 53    & 0104 & 2.1m   & $V$  & $34\times1200$                         & 1.4  & 0.30 &  0.02 &  81.0   & 0.56  &  15.2 &  08 34 08.1 & 66 10 36  \\
                   & 0105 & 2.1m   & $B$  & $25\times1800$                         & 1.6  & 0.30 &  0.02 &             &           &           &                       &                   \\
DDO 86   & 0304 & 2.1m   & $V$  & $25\times1200$                          & 1.3  & 0.30 &  0.03 &  $-$46.0 &  0.86 & 10.6 & 10 44 30.2  & 60 22 04   \\
DDO 133 & 0502 & 4m      & $B$  & $5\times1000$                            & 1.5  & 0.27 &  0.05 & $-$2.0 & 0.69  &  10.0 & 12 32 55.6  &  31 32 16  \\
                  & 0502 & 4m      & $V$  & $10\times900$                           & 1.4   & 0.27 &  0.04 &             &           &           &                       &                    \\
NGC 4163 & 0207 &  2.1m & $V$ & $7\times1800+15\times1200$ &  1.7 & 0.30 &  0.04 &  15.0   &  0.67 &  11.2 &  12 12 09.0  &  36 10 13  \\
                    & 0507 & 2.1m & $B$ & $12\times1800$                          &  1.4 & 0.30 & 0.02 &              &           &           &                        &                   \\
IZw 115   & 0502 & 4m       &  $V$ & $8\times600$                               & 1.1  & 0.27 &  0.04 &  40.7   &  0.47 & 10.0 &  15 32 57.1  &  46 27 10  \\
\enddata
\tablenotetext{a}{Kitt Peak National Observatory telescope used for the observations.}
\tablenotetext{b}{FWHM of a stellar profile on the final combined image.}
\tablenotetext{c}{Photometric calibration rms for the filters listed in Column 4.}
\tablenotetext{d}{Position angle P.A., minor-to-major axis ratio
$b/a$, ellipse semi-major axis step size,
and position of center used to do photometry in concentric ellipses.}
\end{deluxetable}

\clearpage

\begin{deluxetable}{lcccccccc}
\tabletypesize{\small}
\rotate
\tablecaption{Integrated Photometry \label{tab-phot}}
\tablehead{
\colhead{Galaxy}
& \colhead{$M_V$}
& \colhead{$\sigma_{M_V}$}
& \colhead{$B-V$}
& \colhead{$\sigma_{B-V}$}
& \colhead{$M_{NUV}$\tablenotemark{a}}
& \colhead{$\sigma_{M_{NUV}}$}
& \colhead{$FUV-NUV$}
& \colhead{$\sigma_{FUV-NUV}$}
}
\startdata
DDO 53     &  $-$13.80 & 0.01 & 0.23       & 0.01      & $-$12.89 & 0.01 & 0.05 & 0.01 \\
DDO 86     &  $-$16.46 & 0.01 & \nodata & \nodata & $-$15.19 & 0.01 & 0.21 & 0.01 \\
DDO 133   &  $-$15.95 & 0.01 & 0.37       & 0.01      & $-$14.71 & 0.01 & 0.20 & 0.01 \\
NGC 4163 &  $-$14.44 & 0.01 & 0.56       & 0.01      &  $-$12.51 & 0.01 & 0.39 & 0.01 \\
IZw 115      &  $-$16.01 & 0.01 & \nodata & \nodata & $-$14.54 & 0.01 & 0.37 & 0.01 \\
\enddata
\tablenotetext{a}{$M_{NUV}$ is an AB magnitude.}
\end{deluxetable}

\clearpage

\begin{deluxetable}{lccccccccc}
\tabletypesize{\scriptsize}
\rotate
\tablecaption{$V$-band Disk Parameters \label{tab-disk}}
\tablehead{
\colhead{}
& \multicolumn{4}{c}{Inner Exponential}
& \colhead{}
& \multicolumn{4}{c}{Outer Exponential} \\
\cline{2-5} \cline{7-10}
\colhead{}
& \colhead{R$_D$}
& \colhead{$\sigma_{R_D}$}
& \colhead{$\mu_{V,0}$}
& \colhead{$\sigma_{\mu_{V,0}}$}
& \colhead{R$_{Br}$\tablenotemark{a}}
& \colhead{R$_D$}
& \colhead{$\sigma_{R_D}$}
& \colhead{$\mu_{V,0}$}
& \colhead{$\sigma_{\mu_{V,0}}$} \\
\colhead{Galaxy}
& \colhead{(kpc)}
& \colhead{(kpc)}
& \colhead{(mag arcsec$^{-2}$)}
& \colhead{(mag arcsec$^{-2}$)}
& \colhead{(kpc)}
& \colhead{(kpc)}
& \colhead{(kpc)}
& \colhead{(mag arcsec$^{-2}$)}
& \colhead{(mag arcsec$^{-2}$)}
}
\startdata
DDO 53     & 0.60 & 0.03 & 23.74 & 0.13 & 1.98 & 0.43 & 0.01 & 22.36 & 0.20 \\
DDO 86     & 1.72 & 0.09 & 23.18 & 0.10 & 4.70 & 1.16 & 0.07 & 21.76 & 0.34 \\
DDO 133   & 2.16 & 0.09 & 23.80 & 0.05 & 3.80 & 0.69 & 0.01 & 19.70 & 0.14 \\
NGC 4163 & 0.26 & 0.03 & 21.25 & 0.14 & 0.73 & 0.31 & 0.00 & 21.79 & 0.06 \\
IZw 115      & 0.61 & 0.01 & 21.41 & 0.05 & \nodata & \nodata & \nodata & \nodata & \nodata \\
\enddata
\tablenotetext{a}{$R_{Br}$ is the break radius at which the surface brightness profile changes slope.
The profiles of NGC 4163 and IZw 115 do not show this break.}
\end{deluxetable}

\clearpage

\begin{deluxetable}{lccl}
\tablecaption{{\it GALEX} Observations \label{tab-galex}}
\tablehead{
\colhead{}
& \colhead{FUV Exp (s)}
& \colhead{NUV Exp (s)}
& \colhead{Tile Name}
}
\startdata
DDO 53      &   9419  &   8595  & GI4\_015001\_DDO53  \\
DDO 86      & 14050  & 15194  & LOCK\_20 \\
DDO 133    &   9841  & 18287  & GI4\_015003\_DDO133 \\
IZw115       &   8234  &   7616  & GI4\_015005\_IZW115 \\
NGC 4163 & 10173  & 12392  & GI4\_015004\_NGC4163 \\
\enddata
\end{deluxetable}

\clearpage

\begin{deluxetable}{llccr}
\tablecaption{VLA Observations \label{tab-vla}}
\tablehead{
\colhead{} & \colhead{} & \colhead{}
& \colhead{Time on-source} & \colhead{Beam size} \\
\colhead{Galaxy} & \colhead{Date} & \colhead{Array Config}
& \colhead{(hours)} & \colhead{(arcsec)}
}
\startdata
DDO 53   & 1997 Sep \& Nov & CS,D & 3.7,3.6 & 28.5\arcsec$\times$26.7\arcsec \\
DDO 86   & 2006 Dec           & CS   & 4.7       & 15.7\arcsec$\times$13.5\arcsec \\
DDO 133 & 2004 Jun \& Aug   & D    & 8.2   & 48.9\arcsec$\times$45.9\arcsec \\
NGC 4163& 2004 Jun \& Aug  & D    & 7.5   & 50.8\arcsec$\times$46.0\arcsec \\
IZw115       &  2008 Mar             &  CS & 3.4  & 15.1\arcsec$\times$14.6\arcsec  \\
\enddata
\end{deluxetable}

\clearpage

\begin{deluxetable}{lcrrccccc}
\tabletypesize{\scriptsize}
\rotate
\tablecaption{\protect\HI\ Map Parameters \label{tab-himaps}}
\tablehead{
\colhead{} & \colhead{} & \multicolumn{7}{c}{Velocity field fits} \\
\cline{3-9}
\colhead{} & \colhead{log M$_{HI}$} &  \multicolumn{2}{c}{Center (J2000)}
& \colhead{V$_{sys}$} & \colhead{P.A.}
& \colhead{Incl.} & \colhead{max $R$}  & \colhead{Annulii widths} \\
\colhead{Galaxy} & \colhead{(M\protect\solar)} & \colhead{R.A.}
& \colhead{Decl.} & \colhead{(km/s)} & \colhead{(deg)}
& \colhead{(deg)} & \colhead{(arcsec)} & \colhead{(arcsec)}
}
\startdata
DDO 53\tablenotemark{a}     & 7.85 & 08 34 08.9 & 66 11 08  &       17 &      155  & $-304$ to $+79$ & 200 & 25 \\
DDO 86                                     & 9.01 & 10 44 28.8 & 60 22 13  &  1021 &       317 &                         32  & 150 & 15 \\
DDO 133                                   & 8.54 & 12 32 55.6 & 31 32 18  &    331 &       358 &                         46  & 262 &  50 \\
NGC 4163\tablenotemark{b} & 7.20 & 12 12 09.0 & 36 10 13  &    162 &       195 &                         48 &  150 &  25 \\
IZw115                                       & 8.09 & 15 32 56.8 & 46 27 12  &    655 &  50--78 &                         61 &    90 & 15 \\
\enddata
\tablenotetext{a}{Ring parameters are taken from THINGS (Oh et al., in preparation).}
\tablenotetext{b}{Ring parameters are taken from the optical.}
\end{deluxetable}

\clearpage

\begin{deluxetable}{lccccc}
\tabletypesize{\scriptsize}
\rotate
\tablecaption{Star Formation Rates\tablenotemark{a} where  $\mu_V=29.5$ mag arcsec$^{-2}$\label{tab-sfrs}}
\tablehead{
\colhead{} & \colhead{$R$} & \colhead{$\log SFR_{\rm FUV}$} & \colhead{$\sigma_{FUV}$}
& \colhead{$\log SFR_{\rm V}$} & \colhead{$\sigma_{V}$}  \\
\colhead{Galaxy} & \colhead{(kpc)} & \colhead{(M\solar yr$^{-1}$ kpc$^{-2}$)} & \colhead{(M\solar yr$^{-1}$ kpc$^{-2}$)}
& \colhead{(M\solar yr$^{-1}$ kpc$^{-2}$)} & \colhead{(M\solar yr$^{-1}$ kpc$^{-2}$)}
}
\startdata
DDO 53     &  2.8 & -5.5 & 0.8 & -5.1 & 0.1 \\
DDO 86     &  7.5 & -4.7 & 0.3 & -4.8 & 0.2 \\
DDO 133   & 6.1  & -5.1 & 0.5 & -4.9 & 0.1 \\
NGC 4163 &  2.2 & -5.5 & 1.0 & -4.6 & 0.2 \\
IZw115       &  4.6 & -5.0 & 0.6 & -4.9 & 0.2 \\
\enddata
\tablenotetext{a}{$SFR_{\rm FUV}$ is measured from the FUV luminosity.
$SFR_{\rm V}$ is determined from the mass in stars derived from $L_V$. See text for formulae.}
\end{deluxetable}

\end{document}